\documentclass{article}

\usepackage{arxiv}
\usepackage[utf8]{inputenc} 
\usepackage[T1]{fontenc}    
\usepackage{hyperref}       
\usepackage{url}            
\usepackage{booktabs}       
\usepackage{amsfonts}       
\usepackage{nicefrac}       
\usepackage{microtype}      
\usepackage{lipsum}
\usepackage{graphicx}
\usepackage[dvipsnames]{xcolor}
\usepackage{subcaption}
\usepackage{mwe}
\usepackage{amsmath}
\usepackage{multirow}
\usepackage{float}
\usepackage{bm}				
\usepackage{mathrsfs}
\usepackage{verbatim}

\title{Analysing Twitter Semantic Networks: \\ the case of 2018 Italian Elections}

\author{
Tommaso Radicioni\thanks{Corresponding author: tommaso.radicioni@sns.it}\\
Scuola Normale Superiore\\
P.zza dei Cavalieri 7, 56126, Pisa (Italy)
\And
Fabio Saracco
\\
IMT School for Advanced Studies\\
P.zza S. Ponziano 6, 55100, Lucca (Italy)\\
\And
Elena Pavan
\\
University of Trento\\
Via Verdi 26, 38122, Trento (Italy)
\And
Tiziano Squartini
\\
IMT School for Advanced Studies\\
P.zza S. Ponziano 6, 55100, Lucca (Italy)
}

\begin{document}

\flushbottom
\maketitle

\begin{abstract}
Social media play a key role in shaping citizens' political opinion. According to the Eurobarometer, the percentage of EU citizens employing online social networks on a daily basis has increased from 18\% in 2010 to 48\% in 2019. The entwinement between social media and the unfolding of political dynamics has motivated the interest of researchers for the analysis of \emph{users online behavior} - with particular emphasis on \emph{group polarization} during debates and \emph{echo-chambers formation}. In this context, \emph{semantic aspects} have remained largely under-explored. In this paper, we aim at filling this gap by adopting a two-steps approach. First, we identify the \emph{discursive communities} animating the political debate in the run up of the 2018 Italian Elections as groups of users with a significantly-similar retweeting behavior. Second, we study the mechanisms that shape their internal discussions by monitoring, on a daily basis, the structural evolution of the semantic networks they induce. Above and beyond specifying the semantic peculiarities of the Italian electoral competition, our approach innovates studies of online political discussions in two main ways. On the one hand, it grounds semantic analysis within users' behaviors by implementing a method, rooted in statistical theory, that guarantees that our inference of socio-semantic structures is not biased by any unsupported assumption about missing information; on the other, it is completely automated as it does not rest upon any manual labelling (either based on the users' features or on their sharing patterns). These elements make our method applicable to any Twitter discussion regardless of the language or the topic addressed.
\end{abstract}



\section*{Introduction}
\label{sec:introduction}

In the last decade, social media platforms has brought fundamental changes to the way information is produced, communicated, distributed and consumed. According to Eurobarometer, the percentage of Europeans employing online social networks on a daily basis has increased from 18\% in 2010 to 48\% in 2019 \cite{EuroBarometer}. A similar report concerning the US showed that, as of August 2018, 68\% of American adults retrieve at least some of their news on social media \cite{PewResearch.2017}. As social media facilitate rapid information sharing and large-scale information cascades, what emerges is a shift from a \emph{mediated}, \emph{top-down} communication model heavily ruled by legacy media to a \emph{disintermediated}, \emph{horizontal} one in which citizens actively select, share and contribute to the production of politically relevant news and information, in turn affecting the political life of their countries \cite{Schmidt3035}. In a context in which political dynamics unfold with no solution of continuity within a hybrid media space, a multiplicity of studies that cut across traditional disciplinary boundaries have multiplied that uncover the many implications of users online behavior for political participation and democratic processes.

The systematic investigation of online networks spurring from social media use during relevant political events has been particularly helpful in this respect. Endorsing a view of online political activism as complementary to - and not as a substitution for - traditionally studied political participation dynamics \cite{doi:10.1080/14742837.2016.1268956}, detailed and data-intensive explorations of online systems of interactions contributed to a more genuine and multilevel understanding of how social media relate to political participation processes.

At the macro level, research has focused on mapping the structural and processual features of online interaction systems to elaborate on the social media potential for fostering democratic and inclusive political debates. In this respect, specific attention has been paid to assessing grades of polarization and closure \cite{2017NatSR740391D,SCHMIDT20183606} of online discussions within echo-chambers \cite{10.1371/journal.pone.0181821} with a view of connecting such features with the progressive polarization of political dynamics \cite{doi:10.1063/1.4913758, Cherepnalkoski2016}.

At the micro level, attention has gone towards disambiguating the different roles that social media users may play - particularly, to identify influential spreaders \cite{2014NatSR4E5547P,Becatti2019c,Caldarelli2020} responsible for triggering the pervasive diffusion of certain types of information, but also to elaborate on the redefinition of political leadership in comparison to more traditional offline dynamics \cite{doi:10.1177/0002764213479371}. More specifically, accounting for users behavior has helped characterizing the different contributions that are delivered by actors who exploit to different extents social media communication and networking potentials \cite{IJoC1136, doi:10.1111/glob.12119, doi:10.1177/1461444815584764}. In this way, concepts like "political relevance" and "leadership" get redefined at the crossroads between actors' attributes and their actual engagement within online political discussions.

Additionally, increasing attention to online dynamics has entailed dealing with non-human actors, such as platform algorithms \cite{10.5555/2029079} and bots \cite{Cresci2015,Ferrara:2016:RSB:2963119.2818717,Cresci2018,Cresci2019a}. Consideration for non-human actors follows from extant social sciences approaches such as the actor-network theory that invites to disanchor agency from social actors, preferring a recognition for \emph{actants}, i.e. for any agent capable of intervening within social dynamics \cite{Latour2005-LATRTS}. Consistently, the pervasive diffusion of social media in every domain of human action revamps attention for both platform materiality (i.e. the modes in which specific technological artifacts are constructed and function) and for actants, starting from the premise that online dynamics are inherently sociotechnical and, thus, technology features stand in a mutual and co-creative relationship with their social understanding and uses \cite{6734002}. Shrouded in invisibility, platform algorithms and social bots actively filter and/or push specific types of contents thus bending users behaviors and opinions - in some cases acting as true agents of misinformation \cite{doi:10.1080/10584609.2018.1526238, doi:10.1080/17524032.2019.1643383}.

In all its heterogeneity, this variety of studies shares a common feature, insofar as it mostly grounds in the study of \emph{networks of users} and, thus, approaches the study of online political dynamics by privileging the investigation of direct relations amongst actors of different nature - individuals, organizations, institutions and even bots. Conversely, less attention has gone towards the \emph{contents} that circulate during online political discussions and how these contribute to nurture collective political identities which, in turn, drive political action and participation.

To be sure,studies that focus on social media content do exist and embrace a multiplicity of political instances, from electoral campaigns to social movements and protests. For example, looking at Twitter, research has compared the content of tweets published by parties with the content of tweets sent by candidates \cite{doi:10.1177/2056305117704408}, analyzed the contents of the 2017 French presidential electoral campaign \cite{10.1371/journal.pone.0201879}, the online media coverage in the run up of the 2018 Italian Elections \cite{MappingItalianNews} and looked at the keywords and hashtags related to the \textit{\#MeToo} movement \cite{XIONG201910}. Nonetheless, when the focus has been set on social media contents, only rarely have these been investigated in connection with systems of social relations established amongst users on social media platforms~\cite{celli2015long,giglietto2015or}. Ultimately, the social and semantic aspects have hitherto been studied independently and we are still missing empirical pathways to explore the nexus between contents of online political conversations and the relational systems amongst users sustaining them.

This paper aims at filling in this gap by proposing an innovative approach that links the identification of communities of users that display a similar communication behavior with a thorough investigation of the most prominent contents they discuss. More specifically, by looking at the online conversation that unfolded on Twitter in the run up of the 2018 Italian Elections, our paper identifies \emph{discursive communities} as \emph{groups of users with a similar retweeting behavior} and analyses the contents circulating within these online political communities. 


Several techniques of inferring the political alignment of Twitter users have been already implemented looking specifically at electoral campaigns, such as in the run-up to the 2010 U.S. midterm elections \cite{6113114}, the 2012 French presidential elections \cite{10.1145/2508436.2508438} and the 2016 U.S. elections \cite{10.5555/3382225.3382281}. Inference methods adopted in these works are based either on the construction of a retweet network based on the manual labelling of (at least) a fraction of Twitter users or on sharing patterns of web-links and/or hashtags. Moreover, the final number of communities is often defined a-priori, as every link or hashtag is assigned a specific political connotation linked to existing party systems or the left-right ideological spectrum. 

Against this background, the approach we adopt here to infer discursive communities is innovative in two ways. First, it detects groups of Twitter users without any manual labelling (either based on users' features or on content sharing patterns). Second, the political valence of online actions such as retweeting is behavior-driven rather than ideologically determined as it is based on the direct action, performed by unverified users, of expanding the reach of verified users' messages via retweets.


Having identified communities in this way, we turn towards the study of contents they discuss by examining the semantic networks induced by the co-occurrences of hashtags contained in the tweets sent by their members. Through this procedure, it is possible to examine the semantic aspects of online political debates while grounding them in users' behaviors. Finally, we implement several filtering algorithms \cite{Saracco_2017} to detect the non-trivial content of our semantic networks, identifying the most debated subjects. Filtering ultimately allows us to identify the communication strategies adopted by the different discursive communities and the backbone of the narratives developed by the different groups.

The paper is organized as follows. Section \hyperref[sec:data]{Case Study and Data} describes data-acquisition and data-cleaning processes. In Section \hyperref[sec:methods]{Methods}, we discuss the methods we employ to project our bipartite user-hashtag networks on the hashtag layer and to derive our collection of semantic networks. We report and discuss results of our analysis of the Italian case in the Sections \hyperref[sec:results]{Results} and \hyperref[sec:discussion]{Discussion}. Finally, in Section \hyperref[sec:conclusions]{Conclusions} we further elaborate on the potentialities and limits of our proposed approach.

\section*{Case Study and Data}\label{sec:data}

\paragraph{Case study.} The current study focuses on the Twitter discursive communities that emerged during the weeks of the electoral campaign preceding the Italian Elections on 4 March 2018. The 2018 Italian Elections represented a crucial political event that subverted the traditionally bipolar political competition characterizing the so-called Italian Second Republic. A radically novel scenario, with three poles of attraction did in fact emerge. The first pole was represented by the centre-right coalition which eventually won the elections with 37\% of the vote share. Interestingly, the victory of the right-wing alliance was not led by Silvio Berlusconi's party, \textit{Forza Italia} (FI), which obtained only a 14\% of preferences and thus gave way to the nationalist \textit{Lega} led by Matteo Salvini (17,4\%). The second pole was represented by the center-left coalition led by \emph{Partito Democratico} (PD) with 18.7\% of the vote share - its worst result ever - under the leadership of the secretary and Prime Minister candidate Matteo Renzi. The third pole was represented by the populist party \emph{Movimento Cinque Stelle} (M5S) which unexpectedly obtained 32.7\% of preferences.

Ultimately, the 2018 elections constituted a true electoral earthquake triggered by two elements. On the one hand, the extreme predominance of themes such as immigration and criminality which eventually favored populist and right-wing parties over more traditional actors such as \textit{Forza Italia} and the \textit{Partito Democratico}. On the other hand, a significant contribution to the shuffling of political balances was given also by the hybrid electoral campaign \cite{WhosTheWinner} put in place by all leaders and candidates who combined traditional and social media and managed to engage voters with pervasive and low-cost communication strategies.

\paragraph{Social media platform and relations selection.} Twitter is hardly the only social media platform that hosted politically relevant discussions during the observation period, as all social media platforms play an increasingly relevant political role \cite{Bovet2019, DELVICARIO20176}. Nonetheless, extant studies show that Twitter is particularly prominent during electoral dynamics \cite{TwitterElectionCampaigns} as it is the platform used by the vast majority of public figures (e.g. political leaders, journalists, official media accounts, etc.) to provide visibility to their statements. More specifically, in the Italian context, Twitter is recognized to play an "agenda setting" effect on the country's mass media \cite{doi:10.1080/17512786.2015.1040051}. Hence, regardless of the fact that Twitter users are not representative of the Italian population, allows for looking at the discursive communities that emerge on this platform entails looking at a pivotal - albeit non representative - portion of political discussions that accompanied the electoral process.

Amongst all types of interaction modes featured by the platform, the current study grounds on retweets, which we understand as a baseline online relational mechanism that is particularly insightful when studying collective political identities. Indeed, as pointed out in \cite{doi:10.1080/14742837.2016.1268956}, while mentions and replies in Twitter do sustain direct interaction and dialogue between users, retweets suggest a will to re-transmit contents produced by others. This, in turn, provides a more clear-cut indication of commonality and shared points of reference. Moreover, extant research suggests that retweets proxy the actual political alliances better than mentions and replies - as shown in \cite{6113114}, where authors conclude that the use of retweets was more relevant than that of mentions to grasp the bipartisan nature of online debates in the run up of the 2010 US midterm elections.

\paragraph{Data collection.} The extraction of Twitter data has been performed by selecting a set of keywords linked to the Twitter discussion about 2018 Italian Elections. In particular, each collected tweet contains at least one of the following keywords: \textit{elezioni}, \textit{elezioni2018}, \textit{4marzo}, \textit{4marzo2018} (literally, \textit{elections}, \textit{elections2018}, \textit{4march}, \textit{4march2018}). Data collection has been realized by using the Twitter Search API across a period of 51 days - from 28 January 2018 to 19 March 2018 - a time interval covering the entire period of the electoral campaign and the two weeks after the Election Day.

\paragraph{Data cleaning.} The procedure described above led to a data set containing 1.2 millions of tweets, posted by 123.210 users (uniquely identified via their user ID). As in the Twitter environment hashtags play a central role, acting as \emph{thematic tags} designated by the hash symbol \# \cite{doi:10.1080/1369118X.2011.554572}, we defined the nodes of our semantic networks as the hashtags extracted from the text of tweets. Thus, only tweets containing at least one hashtag have been retained. This procedure left us with $\simeq38\%$ of the original dataset. Notably, this result indicates that only a small percentage of users employs at least one hashtag while tweeting, as already reported elsewhere \cite{ICWSM112856}. Hashtags were then subjected to a merging procedure: any two hashtags have been considered as the same if found "similar" and only the most frequent hashtag has been retained. The similarity between hashtags has been quantified through the \textit{Levenshtein} or \textit{edit distance} (see Supplementary Note 1 for more details), i.e. one of the most common sequence-based similarity measures \cite{Survey_TextSimilarity}. This procedure allows us to get rid of hashtag duplicated deriving from typos or linguistic variability. As shown by a check \emph{a posteriori}, our cleaning procedure misidentifies less than $1\%$ of the final list of hashtags.

\paragraph{Data representation.} We then used the lists of user IDs and merged hashtags to define a bipartite network for each day of our observation period (51 bipartite networks in total). A bipartite network is defined by two distinct groups, or layers, of nodes, $\top$ and $\bot$, and only nodes belonging to different layers are allowed to be connected. The bipartite network corresponding to day $t$ can be, thus, represented as a matrix $\mathbf{M}^{(t)}$ whose dimensions are $N_\top\times N_\bot$, with $N_\top$ being the total number of users on day $t$ and $N_\bot$ being the total number of hashtags (tweeted) during that specific day: $m_{i\alpha}^{(t)}=1$ if the user $i$ has tweeted (at least once) the hashtag $\alpha$ on day $t$ and 0 otherwise.

\section*{Methods}\label{sec:methods}

The simplest way to obtain a monopartite projection out of a bipartite network is that of linking any two nodes belonging to the layer of interest (for example, $\alpha$ and $\beta$) if their number of common neighbors is \emph{positive}. Such a procedure yields an $N_\bot\times N_\bot$ adjacency matrix $\mathbf{A}$ whose generic entry reads

\begin{equation}
a_{\alpha\beta}=\Theta[V_{\alpha\beta}^*]
\end{equation}
where 

\begin{equation}
V_{\alpha\beta}^*=\sum_{j=1}^{N_\top}m_{\alpha j}m_{\beta j}
\end{equation}
counts the number of nodes both $\alpha$ and $\beta$ are linked to and $\Theta$ represents the Heaviside step function. The condition $a_{\alpha\beta}=\Theta[V_{\alpha\beta}^*]=1$ can be also rephrased by saying that $\alpha$ and $\beta$ share \emph{at least} one common neighbor.

A more refined method to obtain a monopartite projection is that of linking any two nodes if the number of common neighbors is found to be \emph{statistically significant} according to an entropy-based framework \cite{Saracco_2017}. More quantitatively, this second algorithm prescribes to compare the empirical value $V_{\alpha\beta}^*=\sum_{j=1}^{N_\top}m_{\alpha j}m_{\beta j}$ with the outcome of a properly-defined benchmark model - here, generically indicated with $f$ - via the calculation of the p-value

\begin{equation}
\text{p-value}(V_{\alpha\beta}^{*})=\sum_{V_{\alpha\beta}\geq V_{\alpha\beta}^{*}} f(V_{\alpha\beta})
\end{equation}

and link $\alpha$ and $\beta$ only in case it "survives" a multiple hypotheses test (see Supplementary Note 2 for more details). Such a procedure outputs an $N_\bot\times N_\bot$ adjacency matrix $\mathbf{A}$ whose generic entry reads $a_{\alpha\beta}=1$ if nodes $\alpha$ and $\beta$ are found to be linked to the same neighbors a statistically significant number of times and $a_{\alpha\beta}=0$ otherwise.

The null models used as filters for our analysis are the \emph{Bipartite Random Graph Model} (BiRGM), the \emph{Bipartite Partial Configuration Model} (BiPCM) and the \emph{Bipartite Configuration Model} (BiCM) \cite{Saracco_2017} (see Supplementary Note 2 for more details). In a nutshell, the BiRGM discounts the information provided by the total number of retweets, the BiPCM discounts the information provided by the total number of retweeted hashtags per user, and the BiCM discounts the information provided by both the total number of retweeted hashtags per user and the total number of retweeting users per hashtag.

For every bipartite network in our data set, we created two matrices following the outlined procedure: first, a monopartite user by user network that we employed to identify discursive communities; second, a monopartite hashtag by hashtag network that we employed to study the contents discussed within the identified discursive communities.

\section*{Results}\label{sec:results}

\subsection*{User by user networks and discursive communities}

Our first step to analyze the Twitter public discourse of the 2018 Italian electoral campaign consists of identifying communities of online users with a similar behavior. To this aim, we have divided users into two groups, by distinguishing the accounts \emph{verified} by the platform from the \emph{non-verified} ones. It is worth noting that the account verification procedure can be requested by any user to guarantee to other Twitter users that the account is authentic: for this reason, the verified accounts are usually composed by "entities" such as politicians, journalists, political parties or media. This information can be easily retrieved by employing the Twitter APIs. A bipartite network is, then, built as follows: a verified and a non-verified user are linked if one of the two retweets the other one at least once during the observation period - notably, the retweeting action is mainly performed by non-verified users who share contents published by the verified ones. Then, the procedure described in Section \hyperref[sec:methods]{Methods} has been employed to project the bipartite network of retweets on the layer of verified users employing the BiCM filter. Lastly, a traditional community detection algorithm has been run to identify communities of verified users (see Supplementary Note 3 for more details). These groups constitute \textit{discursive communities} wherein the tweeting activity of the verified users triggers a discussion between the non-verified users sharing similar contents. Moreover, a handful of non-verified users are "assigned" to the communities of the verified ones, via the computation of the so-called \emph{polarization} (see Supplementary Note 4 and also \cite{Becatti2019,2021arXiv210304653R}).

Interestingly, identified discursive communities provide a faithful representation of the alliances running at the 2018 Italian Elections and of their supporters:

\begin{itemize}
\item \textbf{Movimento Cinque Stelle} (\textbf{M5S}): a community composed by accounts of politicians belonging to \textit{Movimento Cinque Stelle} (e.g. \textit{DaniloToninelli}, \textit{luigidimaio}), institutional accounts of the party (e.g. \textit{M5S\_Camera}, \textit{M5S\_Senato}) and users engaging with all of them. The number of users belonging to this community is 11.151;

\item \textbf{Center-right} (CDX): a community of users composed by accounts of allied right-wing political parties (e.g. \textit{forza\_italia}, \textit{LegaSalvini}), center and right-wing politicians (e.g. \textit{renatobrunetta}, \textit{matteosalvinimi}), their institutional representative groups (e.g. \textit{GruppoFICamera}), and users interacting with all of these. The number of users belonging to this community is 5.842;

\item \textbf{Center-left} (CSX): a rather heterogeneous community of users composed by accounts of political parties composing the center-left alliance (e.g. \textit{pdnetwork}, \textit{PD\_ROMA}), their politicians (e.g. \textit{giorgio\_gori}, \textit{matteorenzi}), some journalists (e.g. \textit{vittoriozucconi}, \textit{jacopo\_iacoboni}), and users engaging with them. The number of users belonging to this community is 12.065.
\end{itemize}

\begin{figure}[ht]
\centering
\includegraphics[width=0.8\textwidth]{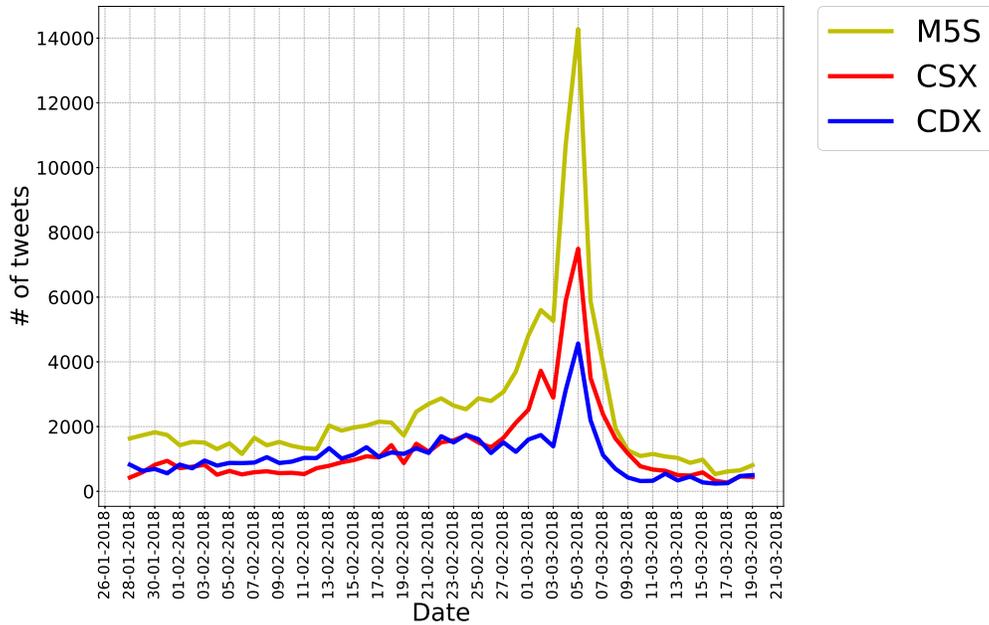}
\caption{Volume of tweets characterizing the M5S, CDX and CSX communities across the observation period: notice the peak of activity, evident for all communities, registered in correspondence of the day \emph{after} elections (5 March 2018). The volume of tweets characterizing the M5S community is systematically larger than the volume of tweets characterizing the other two communities, across the entire period considered here - an element that confirms the attitude of M5S supporters to use social media in political ways.}
\label{fig:1}
\end{figure}

\paragraph{Activity level of discursive communities.} A first step in the analysis of these discursive communities consists of examining their volume of activity. As fig. \ref{fig:1} shows, the evolution of the Twitter activity of the three discursive communities is similar. Generally speaking, a flat trend is followed by a steep rise few days before the Election Day; then, a peak in the tweeting activity is registered the day \emph{after} elections (5 March 2018). Afterwards, a rapid decrease of the number of tweets can be observed: in comparison with values preceding the Election Day, the volume of CDX tweets decreases by $\simeq60\%$, the volume of M5S tweets decreases by $\simeq50\%$ and the volume of CSX tweets decreases by $\simeq20\%$. It is worth noting that the volume of tweets characterizing the M5S community is systematically larger than the volume of tweets characterizing both the CDX and the CSX community across the entire period considered here, an element that confirms the renowned attitude of M5S supporters towards the use of digital media for their organizational and public communication activities.

\subsection*{Hashtag by hashtag networks}

Let us now move to the analysis of the monopartite projections on the layer of hashtags which we label as \emph{hashtag by hashtag} or \emph{semantic networks}. Below, we discuss the results generated by non-filtered projections. In the next section, we compare these results with those generated via filtered projections.

\paragraph{Analyzing the topics prominence.} A closer inspection of semantic networks allows us to engage more systematically with the contents discussed within discursive communities. A first step in this direction can be made by exploring the number of \emph{nodes}, which proxies the number of topics discussed by users, and their \emph{mean degree} (i.e. the mean number of neighbors per node), which proxies the (average) prominence of the topics that characterize the discussion. Results obtained in this step are shown in fig. \ref{fig:2}.

The number of nodes shows a rising trend up to the day \emph{after} elections, followed by a decreasing one. This indicates that the number of topics debated by users increases as 4 March 2018 approaches. Again, the M5S seems to be the most active community with the largest number of debated topics throughout our observation period. The trend characterizing the M5S community is closely followed by that of the right-wing alliance up to the end of February, when an inversion takes place and a rise in the number of topics debated by the supporters of the center-left alliance becomes clearly visible.

The trend of the mean degree is, overall, much less regular: it is, in fact, characterized by several `bumps of activity' throughout the entire period. Peaks in the daily use of hashtags correspond to so-called \textit{mediated events}, i.e., events of social relevance broadcast by mass media and, in particular, by national television channels. The importance of media events is suggested by the prominence of hashtags such as \textit{\#dallavostraparte}, \textit{\#tagadala7}, etc. which all refer to Italian political talk shows. To confirm this intuition, we explored qualitatively the tweets contributing to activity bumps and indeed found a systematic link between contents tweeted by users and mainstream media contents, particularly, television shows featuring prominent politicians. We further cross-referenced the contents of these tweets particularly with printed and online news to verify the actual presence of political leaders and figures during talk shows mentioned by users in their tweet. After this qualitative validation, we concluded that users tend to become particularly active during political debates signalling and/or commenting on the presence of certain candidates in television. Such a behavior is particularly evident for the CDX community, whose mean degree is characterized by a larger number of peaks. More specifically, the peaks are observed in correspondence of the following TV shows:

\begin{figure}[ht]
\centering
\includegraphics[width=\textwidth]{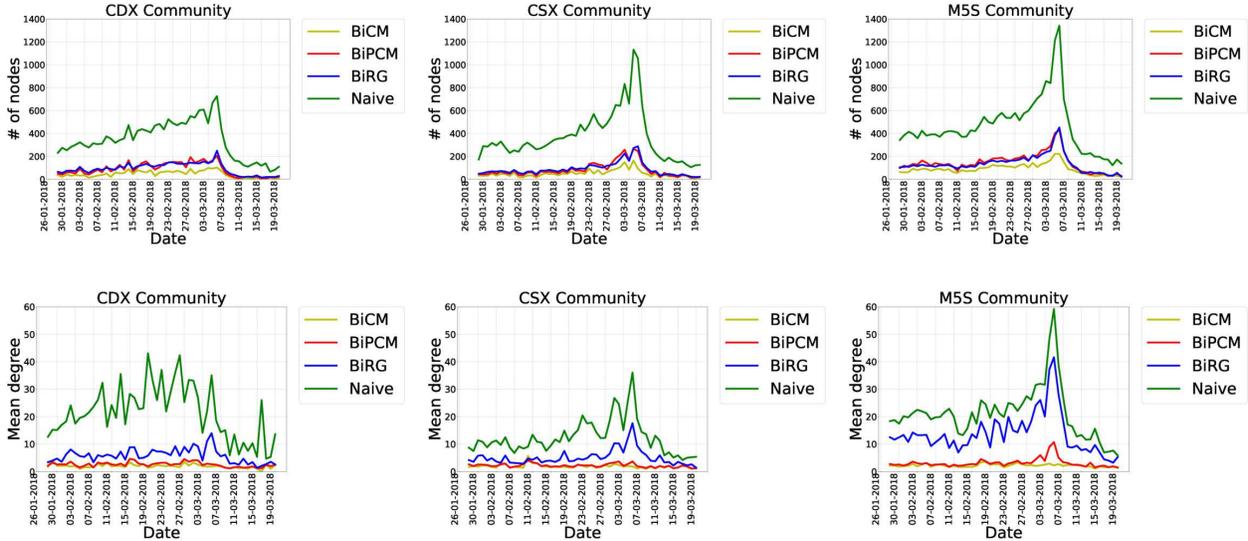}
\caption{Temporal evolution of the number of nodes (top panels) and of the mean degree (bottom panels) for each community-specific semantic network. Peaks in the daily use of hashtags correspond to \emph{mediated events}, i.e. events of social relevance broadcast by media. This behavior is particularly evident for the CDX community whose activity increases in correspondence with TV shows where politicians from the right-wing alliance are hosted.}
\label{fig:2}
\end{figure}

\begin{itemize}
\item \textbf{09 February}: interview of Silvio Berlusconi at \textit{TG La7} (hashtags: \textit{\#silvioberlusconi}, \textit{\#tgla7});
\item \textbf{11 February}: Silvio Berlusconi and Matteo Salvini are interviewed at \textit{Mezz'Ora In Più} (hashtags: \textit{\#il4marzovotaefaivotareforzaitalia}, \textit{\#mezzorainpiù});
\item \textbf{13 February}: Nicola Porro, an italian journalist, announces via a Facebook video, the topics that will be discussed on his TV show \textit{Matrix}, broadcast by \textit{Canale 5}, a TV channel owned by the Berlusconi family (hashtags: \textit{\#nicolaporro}, \textit{\#matrix});
\item \textbf{18 February}: interview of Silvio Berlusconi in the TV show \textit{Che Tempo Che Fa} (hashtags: \textit{\#chetempochefa}, \textit{\#silvioberlusconi});
\item \textbf{19 February}: interview of Silvio Berlusconi in the TV show \textit{Dalla Vostra Parte} (hashtags: \textit{\#dallavostraparte}, \textit{\#silvioberlusconi});
\item \textbf{22 February}: Matteo Salvini and Anna Maria Bernini (from the \textit{Forza Italia} party) are hosted in the TV show \textit{Quinta Colonna} broadcast on \textit{Rete 4}, another TV channel owned by the Berlusconi family (hashtags: \textit{\#forzaitaliaberlusconipresidente}, \textit{\#quintacolonna});
\item \textbf{26 February}: Guido Crosetto and Maurizio Gasparri (both from the right-wing alliance) are hosted in the TV show \textit{L'Aria Che Tira} (hashtag: \textit{\#lariachetirala7});
\item \textbf{16 March}: interview of Michaela Biancofiore (from \textit{Forza Italia}) in the TV show \textit{Tagadà} (hashtags: \textit{\#tagada}, \textit{\#tagadala7}).
\end{itemize}

Beside confirming that Twitter discussions can be \emph{influenced} by external events, our results point out that Twitter discussions can be also \emph{triggered} by external events. This is especially true for the CDX community whose Twitter discussions do not emerge "spontaneously" but are driven by the aforementioned mediated events \cite{10.1371/journal.pone.0094093}, seemingly indicating that CDX users still conceive of television as the main information tool when it comes to political processes.

\begin{table}[ht]
\centering
\begin{tabular}{|c|p{4.5cm}|p{4.5cm}|p{4.5cm}|}
\hline
\hline
$H_t$ & \textbf{M5S} & \textbf{CDX} & \textbf{CSX} \\
\hline
100\% & dimaio, lega, renzi, berlusconi, m5s, pd, italia & salvini, m5s, centrodestra, pd, lega & renzi, salvini, dimaio, m5s, pd \\
\hline
98\% & forzaitalia, salvini & berlusconi, italia, renzi & \\
\hline
96\% & roma, ottoemezzo & forzaitalia & berlusconi, italia, lega \\
\hline
94\% & centrodestra, ricercapubblica & & russia \\
\hline
92\% & boschi, politica & dimaio & europa, politica, roma \\
\hline
90\% & fi, governo & fi, governo & \\
\hline
88\% & casapound & roma & \\
\hline
86\% & meloni & & \\
\hline
84\% & fakenews, lavoro, liberieuguali & casapound, politica & forzaitalia, lavoro, usa \\
\hline
82\% & 8800precari, gentiloni, migranti, senato, voto & governo, lombardia & centrodestra, leu, liberieuguali \\
\hline
80\% & bonino, campagnaelettorale, casini, leu, rosatellum & cdx, flattax, sinistra & milano, partitodemocratico, ue \\
\hline
78\% & avanti, iovotom5s, movimento5stelle, precari, sinistra & lavoro, ue & campagnaelettorale, fakenews, governo \\
\hline
\hline
\end{tabular}
\caption{Hashtag persistence for each discursive community across the entire temporal period covered by our data set (51 days in total). The first column shows the percentage of days each hashtag is present in the set of tweets of each community. Notice that the hashtags that are always present are those carrying the name of political parties and political leaders, while other relevant themes for the political debate are absent from (at least) some of the discursive communities. These findings suggest that the online political debate is largely focused on single personalities/political entities (as particularly evident upon inspecting the CSX hashtags) and only to a much smaller extent on themes of public interest.}
\label{tab:1}
\end{table}

\begin{table}[ht]
\centering
\begin{tabular}{|c|p{4.5cm}|p{4.5cm}|p{4.5cm}|}
\hline
\hline
$T_t$ & \textbf{M5S} & \textbf{CDX} & \textbf{CSX} \\
\hline
31\% & (ricercapubblica, 8800precari, campagnaelettorale) & & \\
\hline
27\% & & & (salvini, pd, m5s) \\
\hline
24\% & (pd, italia, m5s) & & (pd, lega, m5s); (pd, dimaio, m5s) \\
\hline
21\% & (cnr, campagnaelettorale, ricercapubblica); (precari, campagnaelettorale, ricercapubblica); (politica, pd, m5s); (dimaio, pd, m5s); (lega, pd, m5s)  & & (m5s, dimaio, salvini); (liberieuguali, pd,m5s); (m5s, berlusconi, pd); (usa, europa, russia); (savona, accettolasfida, poterealpopolo) \\
\hline
20\% & (berlusconi, pd, m5s); (ottoemezzo, pd, m5s); (salvini, pd, m5s); (berlusconi, politica, m5s); (centrodestra, pd, m5s); (italia, stopinvasione, italiani); (italia, stopislam, italiani); (campagnaelettorale, piemonte, forzaitalia); (m5s, pd, m5salgoverno) & (salvini, pd, m5s) & (pd, m5s, renzi); (pd, italia, m5s); (salvini, lega, m5s); (forzaitalia, pd, m5s); (fattinonparole, partitodemocratico, avanti); (berlusconi, salvini, pd); (salvini, m5s, berlusconi) \\
\hline
\hline
& \multicolumn{3}{|c|}{\textbf{Dates}} \\
\hline
& 02 Mar 2018 & 20 Feb 2018 & 02 Mar 2018 \\
& 20 Feb 2018 & 27 Feb 2018 & 06 Mar 2018 \\
& 21 Feb 2018 & 02 Mar 2018 & 23 Feb 2018 \\
& 07 Mar 2018 & 01 Mar 2018 & 04 Mar 2018 \\
& 16 Feb 2018 & 22 Feb 2018 & 21 Feb 2018 \\
\hline
\hline
\end{tabular}
\caption{Persistence of triadic closures for each discursive community across the entire temporal period covered by our data set (51 days in total), on non-filtered projections. This analysis is particularly insightful to distinguish the behavior of the three communities: while the CSX and CDX communities are characterized by triads exclusively about political leaders, parties and electoral slogans, the triads observed within the M5S community focus more on themes of public interest. Notice that the largest $T_t$ value, i.e. the largest percentage of days a specific triadic closure is present in our data set, is sensibly less than the number of days covered by our data set (i.e. 51). Dates refer to the days with the largest number of triadic closures.}
\label{tab:2}
\end{table}

\paragraph{Identifying persistent topics.} A second step towards a closer understanding of the contents discussed within discursive communities consists of quantifying the \emph{interest towards a topic} throughout the entire period covered by our data set. To this aim, we analyzed \emph{hashtag persistence}, $H_{t}$, i.e. the percentage of days an hashtag is present in our data set. Results are reported in table \ref{tab:1}. As it shows, the most persistent hashtags (in fact, the ones that are always present) are those concerning the name of political parties (i.e. \textit{\#lega}, \textit{\#m5s}, \textit{\#pd}) and political leaders (i.e. \textit{\#berlusconi}, \textit{\#dimaio}, \textit{\#renzi}, \textit{\#salvini}). Moreover, more persistent hashtags in all discursive communities refer almost in all cases to political actors and figures, more often than not of an opposing alliance. When it comes to substantive electoral themes, instead, the three communities seem to hold a common interest for work-related matters but also concentrate on peculiar interests: migration flows for the M5S, taxation for the CDX and the role of Europe for the CSX. This finding has been observed for all discursive communities and it highlights the fact that the online political debate largely focuses on single personalities/political entities and, albeit to a lesser extent, on themes of public interest.

\paragraph{Identifying central topics.} In order to identify topics that, regardless of their prominence and persistence, are more pivotal to the unfolding of the discussion, we computed \emph{hashtag betweenness centrality}, a measure quantifying the percentage of shortest paths passing through each hashtag, i.e.

\begin{equation}
b_\gamma=\sum_{\beta(\neq\alpha)}\sum_\alpha\frac{\sigma^{\alpha\beta}_\gamma}{\sigma^{\alpha\beta}}
\end{equation}

(where $\sigma^{\alpha\beta}_\gamma$ is the number of shortest paths between hashtags $\alpha$ and $\beta$ passing through hashtag $\gamma$ and $\sigma^{\alpha\beta}$ is the total number of shortest paths between hashtags $\alpha$ and $\beta$). In this sense, hashtag betweenness centrality provides an entry point to identify strategic topics that "coordinate" the discussion, as they bridge other topics that users do not directly connect within their tweets. Interestingly, the basket of the most strategic hashtags (i.e. \textit{\#pd}, \textit{\#m5s}, \textit{\#renzi}, \textit{\#salvini}, \textit{\#berlusconi}, \textit{\#italia}, \textit{\#dimaio}, \textit{\#lega}, \textit{\#centrodestra}) is basically the same for all communities. This result reveals how the overall discussion is highly personalized as the main players of the 2018 Italian Elections embody crucial concepts for the definition of the narratives shaping the political debates of \emph{all} communities. Nonetheless, the specificities of each community are maintained when it comes to economic and societal issues.

\paragraph{Analysis of triadic closures.} As discussions develop around communities of topics, increasingly complex semantic structures are to be considered. To this aim, we analyzed the presence and the persistence of \emph{triadic closures}, i.e. \emph{triangles of connected hashtags}, which in our approach represent the core of (larger) semantic networks. In other words, they are the "seeds" around which more complex discussions grow - exactly as triangular motifs represent the simplest (yet informative) example of communities~\cite{2014PhRvE.90d2806B}.

As it has been noticed, this kind of structure provides deeper insights into users' tweeting behavior, by revealing which concepts appear \emph{simultaneously} in a discussion and measuring how often they do so \cite{David:2010:NCM:1805895}. This analysis is particularly insightful to distinguish the behavior of the three communities: as shown in table \ref{tab:2}, while both the CDX and the CSX communities are characterized by triads of concepts exclusively about political leaders, parties and electoral slogans, the triads observed within the M5S community reveal a greater concern of their members for themes of public interest (e.g. the issues of precarious labour, migrants landing, public research).

Interestingly, we also notice that specific days exist in which an abundance of triadic closures is registered. For instance, on the first day of the electoral silence, i.e. 2 March 2018, users are particularly active in building narratives around electoral slogans, while themes of public interest constitute the topic of tweets at the end of the electoral campaign (i.e. the last days of February). Finally, we notice that the abundance of hashtag triads tends to rise in correspondence with mediated events, as observed for the mean degree: this is the case for the days 27 February 2018 for the M5S community (when Luigi Di Maio was interviewed at the political talk show \textit{diMarted\`i}), 20 February 2018 for the CDX community (when Silvio Berlusconi was interviewed in a talk show called \textit{\#Italia18} organized by the Italian newspaper \emph{Corriere della Sera}) and 23 February 2018 for the CSX community (when Laura Boldrini was interviewed at the radio show \textit{Circo Massimo}).

\begin{figure}[ht]
\includegraphics[width=\linewidth]{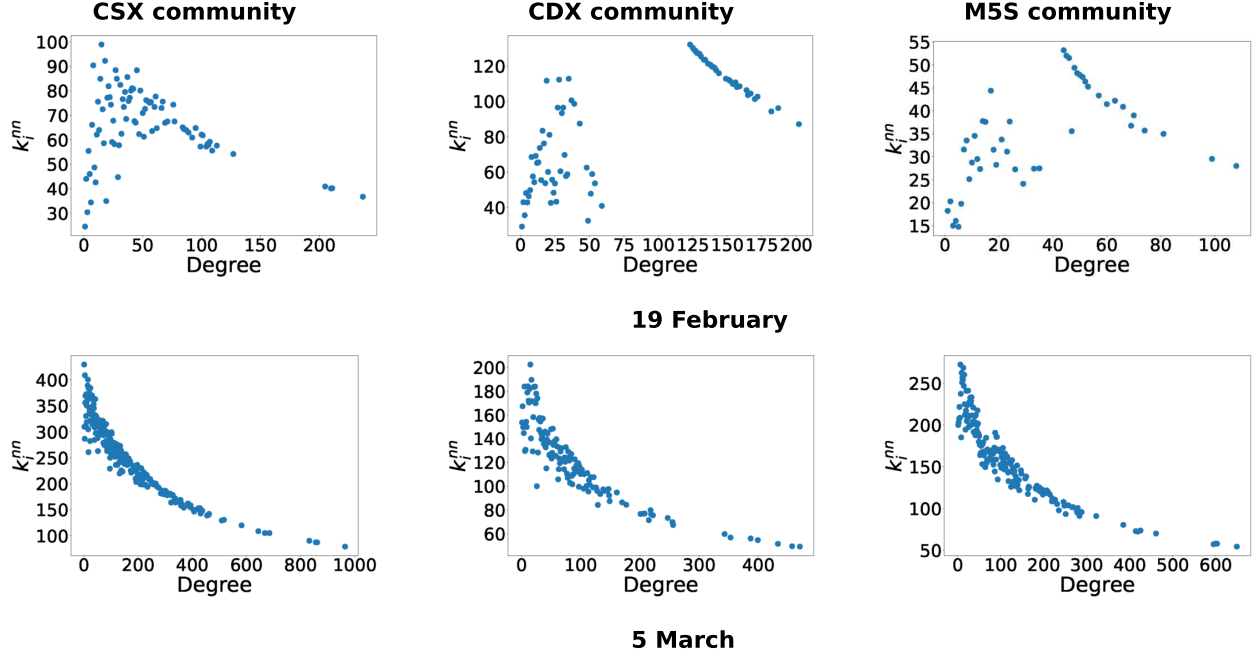}
\caption{Analysis of the degree-degree correlations for two specific days, i.e. 19 February and 05 March 2018. As the trend of $\kappa_\alpha^{nn}$ reveals, the daily semantic networks are disassortative for all communities, i.e. nodes with small degree are (preferentially) connected to nodes with high degree and vice versa. A close inspection of behaviors of the CDX and the CSX communities shows groups of nodes with a (much) larger value of the ANND: these clusters of hashtags constitute the core of the Twitter discussion in the corresponding community which appear in correspondence of specific events and disappear the day after.}
\label{fig:3}
\end{figure}

\paragraph{Analysis of degree-degree correlations.} A closer inspection of correlations between hashtags degrees allows us to elaborate more in depth on the ways prominent topics are connected to others, shaping broader politically relevant narratives. To this aim, we consider the \textit{average nearest-neighbors degree} (ANND), defined, for the generic hashtag $\alpha$, as the arithmetic mean of the degrees of the neighbors of a node, i.e.

\begin{equation}
\kappa_{\alpha}^{nn}=\frac{\sum_{\beta(\ne\alpha)}a_{\alpha\beta}\kappa_{\beta}}{\kappa_{\alpha}},\quad\forall\:\alpha
\end{equation}

with $\kappa_\alpha$ indicating the degree of the hashtag $\alpha$ in the considered monopartite projection. The degree-degree correlation structure of a network can be easily inspected by plotting the $\kappa_{\alpha}^{nn}$ values versus the $\kappa_\alpha$ values. A \emph{decreasing trend} would lead to conclude that correlations between degrees are \emph{negative} - that is, nodes with a small degree would be "preferentially" connected to nodes with high degree and vice versa. Conversely, an \emph{increasing trend} would signal that correlations between nodes are \emph{positive} - that is, nodes with a small (large) degree would be "preferentially" connected to nodes with a small (large) degree. Thus, decreasing and increasing trends offer us an entry point to explore whether discussions in the three communities tend to anchor to some key themes that work as conversational drivers.

The decreasing behavior of the ANND throughout our data set confirms the presence of negative degree-degree correlations, i.e. the considered networks are \emph{disassortative} (less prominent hashtags are connected with more prominent hashtags and vice versa). Examples of these trends are reported in fig. \ref{fig:3}. The days considered here, i.e. 19 February 2018 and 5 March 2018, have been chosen to highlight an interesting feature of our semantic networks: as it is clearly visible upon inspecting the behavior of the CDX and the CSX communities, groups of nodes with a (much) larger value of the ANND appear. As it will become evident in what follows, these hashtags constitute the \emph{core} of the Twitter discussion in the corresponding community and are characterized by a dynamics on a daily time-scale, i.e. they appear in correspondence of a specific event (in the case of the CDX community, the interview of Silvio Berlusconi in a TV show; in the case of the CSX community, Laura Boldrini's Twitter campaign) and disappear the day after.

\begin{figure}[ht]
\includegraphics[width=\linewidth]{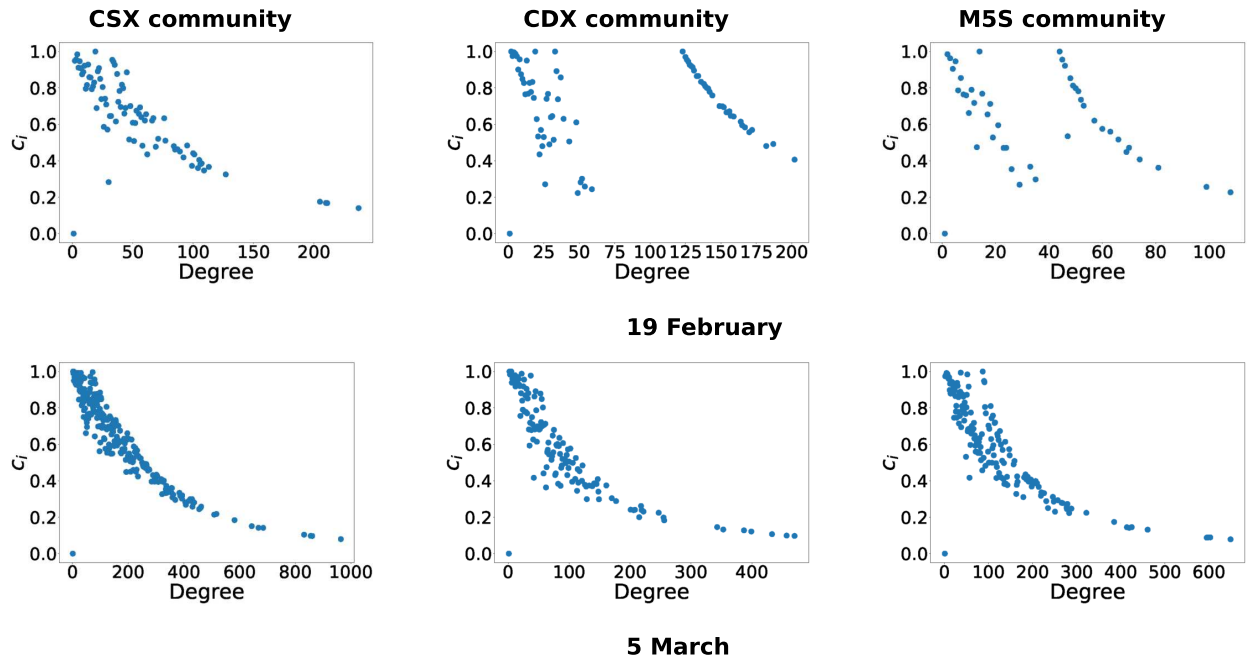}
\caption{Analysis of network hierarchical structure for two specific days, i.e. 19 February 2018 and 5 March 2018. Plotting the clustering coefficient $c_\alpha$ values versus the degree $\kappa_\alpha$ values for the three communities reveals that our daily semantic networks are hierarchical, i.e. poorly-connected hashtags are strongly interconnected and vice versa. Besides, it also shows that the nodes with a larger value of the ANND are also the ones characterized by a larger value of the clustering coefficient.}
\label{fig:4}
\end{figure}

As an additional analysis, we also considered the \textit{clustering coefficient}, defined as

\begin{equation}
c_\alpha=\frac{\sum_{\gamma(\ne\alpha,\beta)}\sum_{\beta(\ne\alpha)}a_{\alpha\beta}a_{\beta\gamma}a_{\gamma\alpha}}{\kappa_{\alpha}(\kappa_{\alpha}-1)},\quad\forall\:\alpha
\end{equation}

and quantifying the percentage of neighbours of a given node $\alpha$ that are also neighbours of each other (i.e. the percentage of triangles, having $\alpha$ as a vertex, that are actually present). As shown in fig. \ref{fig:4}, decreasing trends are observed: poorly-connected hashtags are strongly inter-connected and vice versa, thus suggesting the presence of several inter-connected "small" discussions that are connected to a set of central topics. A network with these features is also said to be \emph{hierarchical}. Moreover, the hashtags with a larger value of ANND are also those with a larger value of the clustering coefficient - confirming the "coreness" of this group of topics. Taken altogether, these results suggest that all discursive communities revolve around a handful of few conversational drivers: overshadowed by the predominance of these issues, a set of niche discussions tend nonetheless to emerge, pointing out a variety of interests even within a single discursive community.  

\paragraph{Semantic networks at the mesoscale: k-core decomposition.} Shifting perspective onto the mesoscale structure of semantic network helps us to better clarify the power of conversational drivers we just identified. If triadic closure allowed us to identify the seeds to thematic discussions within communities, broadening the scope of the semantic analysis to the level of clusters allows us to explore the main conversational lines within the different communities.

In what follows, we focus our attention on 19 February 2018, but similar considerations hold true for other daily semantic networks. We implement the so-called \emph{k-core decomposition}, a technique which has been widely used to find the structural properties of networks across a broad range of disciplines including ecology, economics and social sciences \cite{KONG20191}. The k-core decomposition can be described as a sort of pruning process, where the nodes that have degree less than $k$ are removed, in order to identify the largest subgraph of a network whose nodes have \emph{at least} $k$ neighbors. This method allows a "coreness" score to be assigned to each node of the network which remains naturally divided into shells. Node coreness is equal to $k$ if the node is present in the $k$-core of the network but not in the $(k+1)$-core.

\begin{figure}[ht]
\centering
\includegraphics[width=\textwidth]{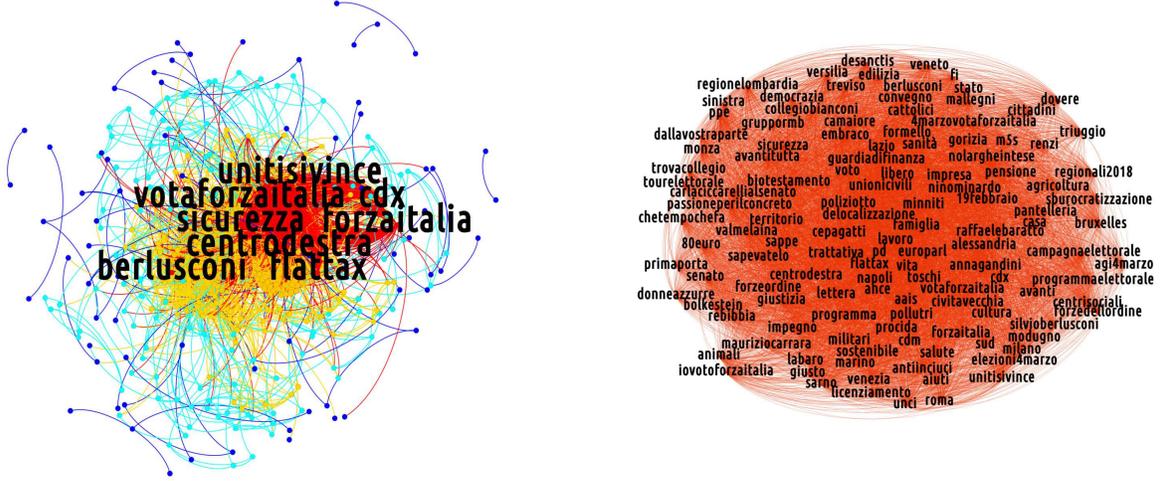}
\caption{K-core decomposition of the semantic network for the non-filtered projection of the CDX discursive community on 19 February 2018. In the left plot, five k-shells are represented with different colors. In the right plot, an expanded view of the innermost k-shell - basically overlapping with the properly defined core individuated by the bimodular surprise - is represented. The compact bulk is triggered by the interview of Silvio Berlusconi in the TV show \textit{Dalla Vostra Parte}. [Picture realized with Gephi software version 0.9.2. For more information about Gephi \cite{ICWSM09154}: \url{https://gephi.org/}]}
\label{fig:5}
\end{figure}

\begin{figure}[ht]
\centering
\includegraphics[width=\textwidth]{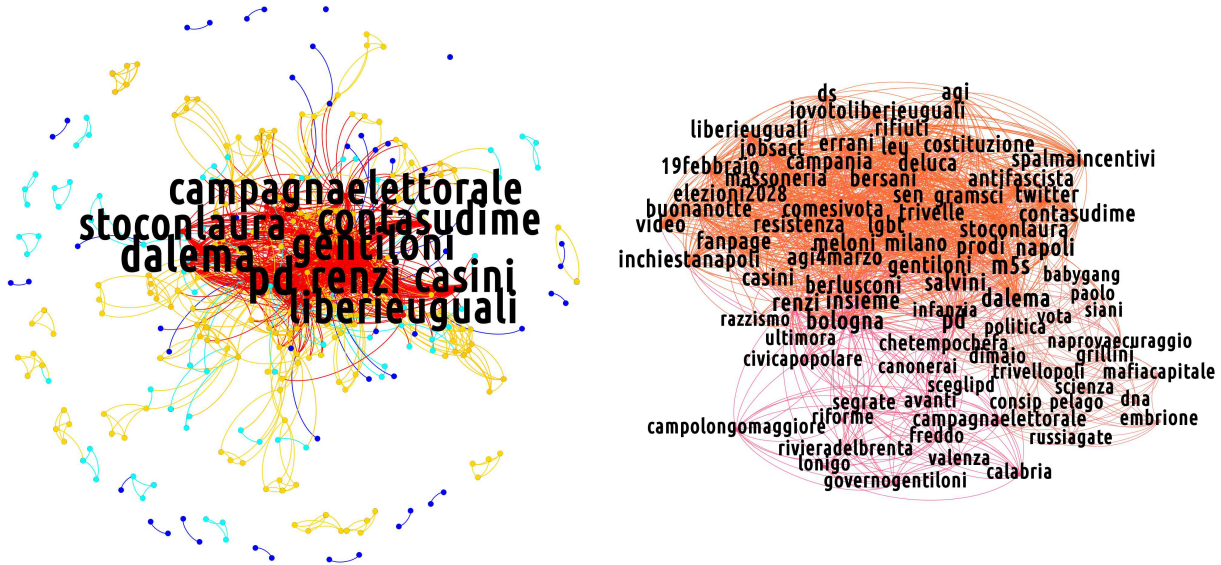}
\caption{K-core decomposition of the semantic network for the non-filtered projection of the CSX discursive community on 19 February 2018. In the left plot, five k-shells are represented with different colors. In the right plot, an expanded view of the innermost k-shell - basically overlapping with the properly defined core identified by the bimodular surprise - is represented. Notice the presence of communities, found through the Louvain algorithm and emerging as a consequence of factors as diverse as the Twitter campaign born in support of the center-left candidate Laura Boldrini, the visit of Matteo Renzi in Bologna, the presence of Massimo D'Alema (another leader of the center-left alliance) in the radio show \textit{Circo Massimo}. [Picture realized with Gephi software version 0.9.2. For more information about Gephi \cite{ICWSM09154}: \url{https://gephi.org/}]}
\label{fig:6}
\end{figure}

\begin{figure}[ht]
\centering
\includegraphics[width=\textwidth]{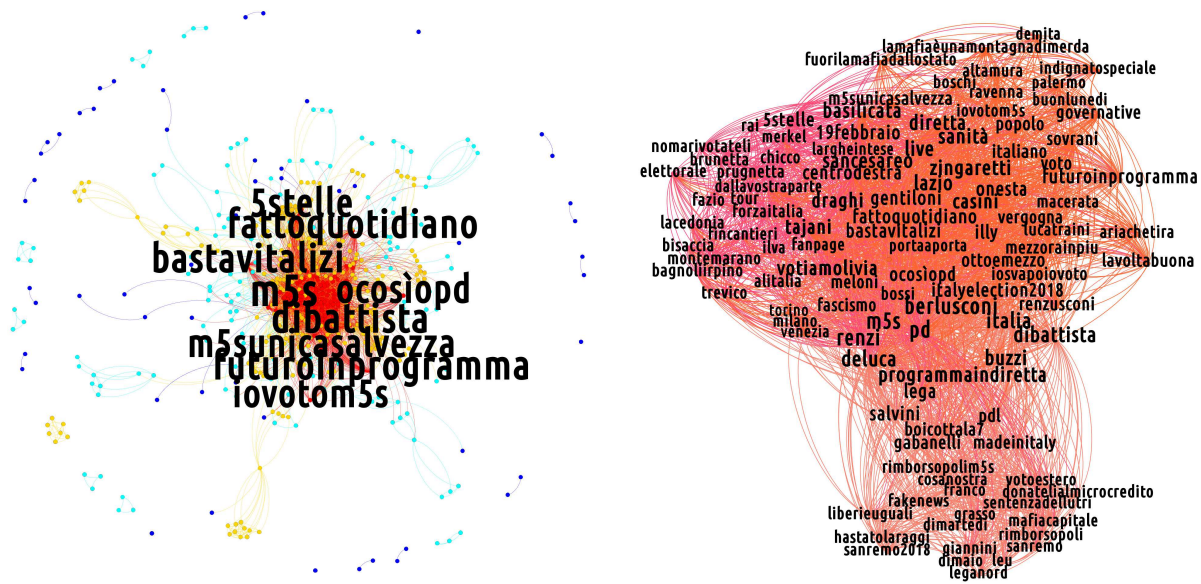}
\caption{K-core decomposition of the semantic network for the non-filtered projection of the M5S discursive community on 19 February 2018. In the left plot, five k-shells are represented with different colors. In the right plot, an expanded view of the innermost k-shell - basically overlapping with the properly defined core identified by the bimodular surprise - is represented. Notice the presence of communities, found by running the Louvain algorithm. These emerge as a consequence of events as the electoral tour of Alessandro Di Battista (one of the M5S leaders), the presence of politicians in TV shows such as \textit{Porta a Porta}, \textit{Mezz'Ora In Più} and \textit{Dalla Vostra Parte}. [Picture realized with Gephi software version 0.9.2. For more information about Gephi \cite{ICWSM09154}: \url{https://gephi.org/}]}
\label{fig:7}
\end{figure}

Figures \ref{fig:5}, \ref{fig:6}, \ref{fig:7} show the the k-shell decomposition for the semantic networks of our discursive communities, for the day 19 February 2018: five k-shells, corresponding to five quantiles of the degree distribution, have been colored, confirming the presence of a core of highly debated hashtags (the red one collecting the most prominent and intertwined ones). To inspect the presence of a substructure nested into the discussion core, we run the Louvain algorithm on the innermost k-shell of the semantic networks of our discursive communities. Their shell structure is indeed rich, as particularly evident when considering the CSX and the M5S ones: indeed, several communities appear, seemingly indicating that the discussions in which members of the two groups are (more) engaged self-organize around sub-topics.

For what concerns the CSX community, the hashtags sub-communities emerge as a consequence of factors as the Twitter campaign born in support of the center-left candidate Laura Boldrini (revealed by the presence of hashtags such as \textit{\#stoconlaura} and \textit{\#contasudime}), the visit of Matteo Renzi in Bologna (revealed by the presence of hashtags such as \textit{\#bologna}, \textit{\#renzi}, \textit{\#errani}, \textit{\#casini}, hashtags that refer to Vasco Errani and Pier Ferdinando Casini, center-left wing candidates for the Senate in Emilia-Romagna) and the presence of Massimo D'Alema (another leader of the center-left alliance) in the radio show \textit{Circo Massimo} (revealed by the presence of hashtags such as \textit{\#dalema}).

On the other hand, the presence of multiple debates within the core of the M5S semantic network relates to events like the electoral tour of Alessandro Di Battista, a prominent figure of the party, who presented the M5S electoral program in the southern Italy region named Basilicata (hashtags: \textit{\#dibattista}, \textit{\#ilfuturoinprogramma}, \textit{\#programmaindiretta}, \textit{\#basilicata}), the presence of a journalist of \textit{Il Fatto Quotidiano} (a newspaper supporting the M5S) invited in the TV show \textit{Otto e Mezzo} (hashtags: \textit{\#ilfattoquotidiano}, \textit{\#ottoemezzo}) and the presence of politicians supporting other coalitions in several TV shows such as \textit{Porta A Porta}, \textit{Mezz'Ora In Più} and \textit{Dalla Vostra Parte}.

However, these observations do not hold true when the CDX-induced semantic network is considered. Its innermost shell is, in fact, a compact group of topics that cannot be further partitioned. 

As a second observation, we notice that - when present - the communities partitioning the core are "held together" by the nodes with largest betweenness centrality: as they coincide with the hashtags related to the name of political parties/leaders, the latter ones can be imagined to act as bridges connecting different discussions. Generally speaking, this indicates that the concept of "most influential nodes" can be applied also within the core of the networks of hashtags - a result that complements the one about the influential spreaders individuated within the networks of users \cite{Kitsak2010}.

\paragraph{Semantic networks at the mesoscale: the core-periphery structure.} In order to complement the analysis above, we also implemented the method proposed in \cite{de2019detecting} which prescribe to search for the network core-periphery partition minimizing the quantity called \emph{bimodular surprise}, i.e.

\begin{equation}
\mathscr{S}_\parallel=\sum_{i\geq l_\bullet^*}\sum_{j\geq l_\circ^*}\frac{\binom{V_\bullet}{i}\binom{V_\circ}{j}\binom{V-(V_\bullet+V_\circ)}{L-(i+j)}}{\binom{V}{L}}.
\label{eq2}
\end{equation}

The quantity above is the multinomial version of the \emph{surprise}, originally proposed to carry out a \emph{community detection} exercise \cite{de2019detecting}. In our case, $L$ is the total number of links observed in our projections, while $V$ is the total number of possible links, i.e. $V=\frac{N(N-1)}{2}$. The quantities marked with $\bullet$ ($\circ$) refer to the corresponding core (periphery) quantities: for example, $l_\bullet^*$ is the number of observed links within the core, while $l_\circ^*$ is the number of observed links within the periphery. The presence of three binomial coefficients allows three different "species" of links to be accounted for: the binomial coefficient $\binom{V_\bullet}{i}$ enumerates the number of ways $i$ links can redistributed \emph{within} the core, the binomial coefficient $\binom{V_\circ}{j}$ enumerates the number of ways $j$ links can redistributed \emph{within} the periphery, and the binomial coefficient $\binom{V-(V_\bullet+V_\circ)}{L-(i+j)}$ enumerates the number of ways the remaining $L-(i+j)$ links can be redistributed \emph{between} the two, i.e. over the remaining $V-(V_\bullet+V_\circ)$ node pairs (see Supplementary Note 3 for more details).

The mesoscale structure characterizing all discursive communities consists of a group of very well-connected vertices linked to a group of low-degree, loosely inter-linked nodes, see figs. \ref{fig:5}, \ref{fig:6}, \ref{fig:7}. Such a structure is known as \emph{core-periphery} and is present in many social, economic and financial systems~\cite{malvestio2020interplay}. Remarkably, nodes belonging to the innermost shell overlap with the core ones computed with the multinomial version of the \emph{surprise}, as proved by computing the \emph{Jaccard index} over two sets of nodes which is a measure of similarity between two sets of elements and is defined as the size of the intersection divided by the size of the union of the two sets: $J(A,B)=\frac{|A\cap B|}{|A\cup B|}$.

\begin{figure}[ht]
\centering
\includegraphics[width=0.8\textwidth]{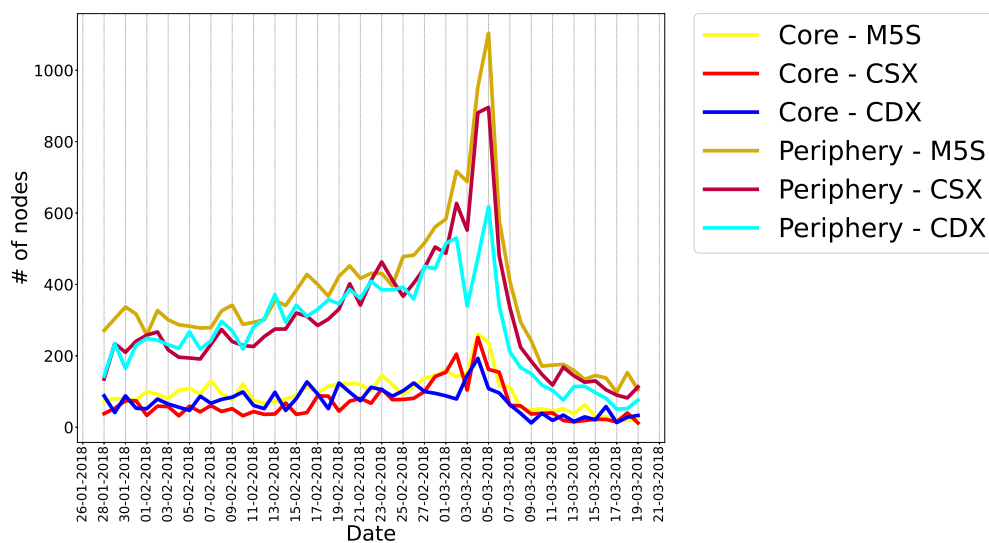}
\caption{Evolution of the number of nodes belonging to the core and to the periphery of each discursive community. The core size is nearly constant throughout the observation period while the periphery size rises as the Election Day approaches (the peak appears in correspondence of the day after, i.e. the 5 March 2018). This behavior, common to all communities,reveals that, as the Election Day approaches, the number of topics animating the discussion increases.}
\label{fig:8}
\end{figure}

As a last comment, let us explicitly show the evolution of the number of nodes belonging to the core and to the periphery for each discursive community. As fig. \ref{fig:8} shows, the core size is nearly constant throughout all the considered period while the periphery size rises in correspondence of Election Day, showing a peak in correspondence of the day after the elections (i.e. 5 March 2018). This behavior, common to all communities, seems to indicate that, as the Election Day approaches, the number of topics discussed does increase.


\subsection*{Filtering the projection}

Let us now focus on the structural features of the filtered projections. Before presenting the results of the analysis, we briefly recall how the filtering procedure works.

Filtering lets the statistically significant overlaps of hashtags measured on the real system emerge. More in detail, for any couple of hashtags, we count how many users are employing them. Then, we consider as a benchmark a set of networks that preserve on average some information of the user-hashtag bipartite network, i.e. the total number of links (as in a bipartite Erd\"os-R\'enyi, or Bipartite Random Graph, \emph{BiRGM}), the degree sequence of the hashtag layer (Bipartite Partial Configuration Model, \emph{BiPCM}) or the degree sequence of both layers (Bipartite Configuration Model, \emph{BiCM}). Naturally, the stricter the constraints (i.e. the
properties preserved by the ensemble), the more detailed the description of the ensemble, as compared to the real network, and fewer edges are validated. In this sense, the number
of validated co-occurences are those non compatible with the expected structural features of the different null models. In other words, the validation procedure retains the non-trivial
co-occurences of hashtags present in the initial bipartite structure, namely those that are not explained by the constraints of the null model used for filtering. Among the aforementioned null models, the BiRGM is expected to retain the highest number of edges between hashtags while the BiCM is the most restrictive among the filtering null models.

Such a procedure has been implemented in previous studies to detect the backbone of the network structure, filtering the real system from random noise, and to highlight non trivial behaviors in the original system \cite{Saracco_2017,Straka2017}. In the present case, the filtering procedure allows the detection not only of \emph{extremely popular hashtags} but also of those hashtags connected only to a single message retweeted by a significant
number of users. The former, in fact, are single hashtags tweeted by a huge amount of users and therefore having a large number of co-occurences, a feature that is compatible with at
least one of the null models considered here and thus possibly filtered out by this procedure. The latter, instead, are more likely to be groups of hashtags whose non-trivial co-occurrence
will survive the filtering procedure. As an example, a set of hashtags used together only in one tweet can be considered: if
this particular tweet is retweeted a significantly high number of times, the number of co-occurences of these hashtags appearing once is given by the number of times the original
message has been retweeted plus the contribution of the original message. In this case, as the null model distribute the co-occurence probability among all the other hashtags in the semantic network, the probability for their co-occurence to be validated thus becomes larger. Differently, a popular hashtag published by a large number of users has a higher probability to co-occur with other hashtags and, at which point the null model could explain its co-occurrence with these hashtags as induced by the network properties. In other words,
the filtering algorithm presented in this work is more likely to discard co-occurences with popular hashtags than with these hashtags appearing together in a single tweet shared a significant number of times. 

\begin{figure}[ht!]
\centering
\includegraphics[width=\textwidth]{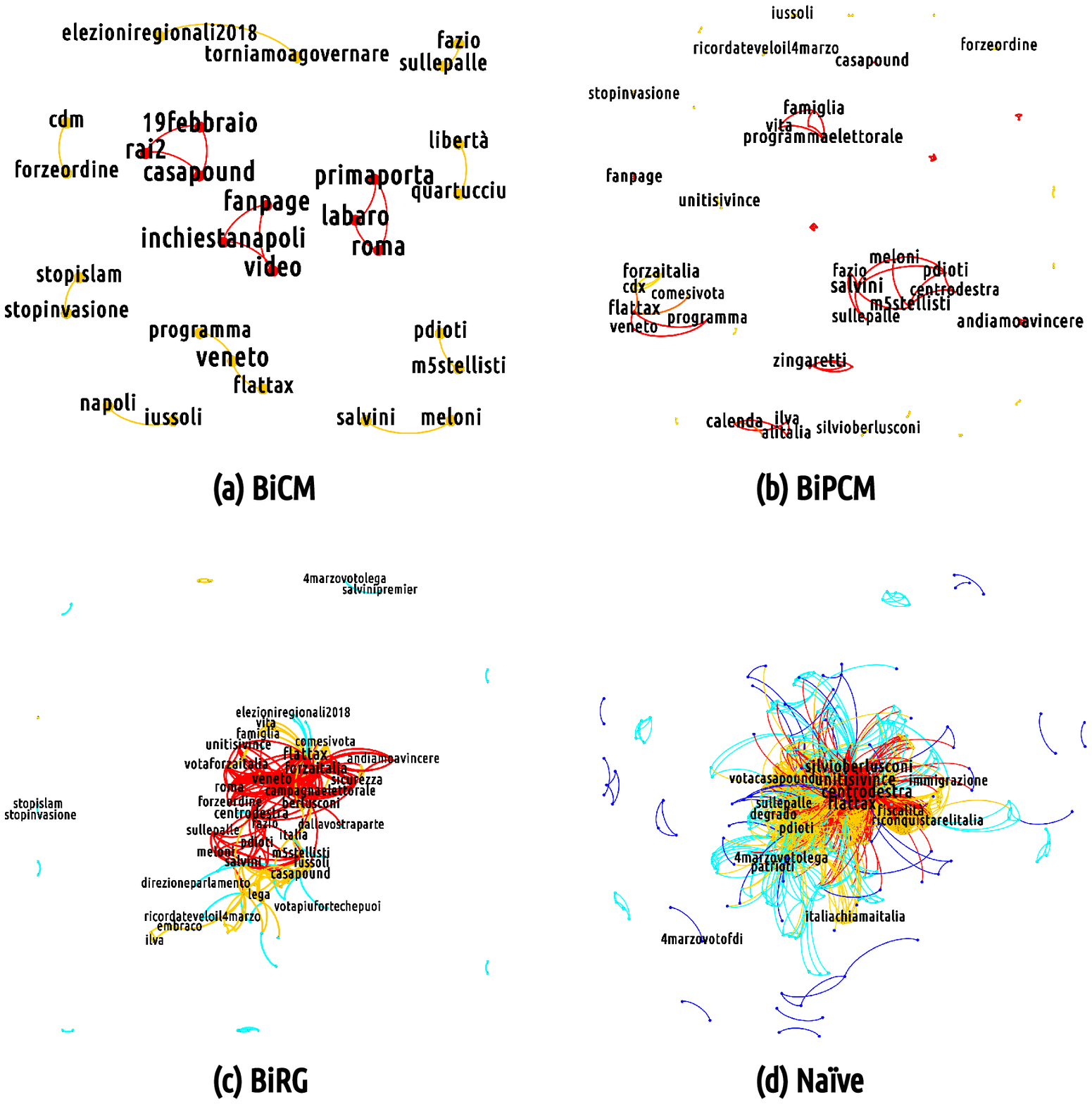}
\caption{Mesoscale structure of (from bottom-right, clockwise) the non-filtered projection of the semantic network corresponding to the CDX discursive community on 19 February 2018 and of the projection of the same network filtered according to the BiRGM, the BiCM and the BiPCM, respectively. The BiCM lets only \emph{few} hashtags survive, reading \textit{\#iussoli}, \textit{\#sicurezza}, \textit{\#stopinvasione}, \textit{\#stopislam}. [Picture realized with Gephi software version 0.9.2. For more information about Gephi \cite{ICWSM09154}: \url{https://gephi.org/}]}
\label{fig:9}
\end{figure}

\begin{figure}[ht!]
\centering
\includegraphics[width=\textwidth]{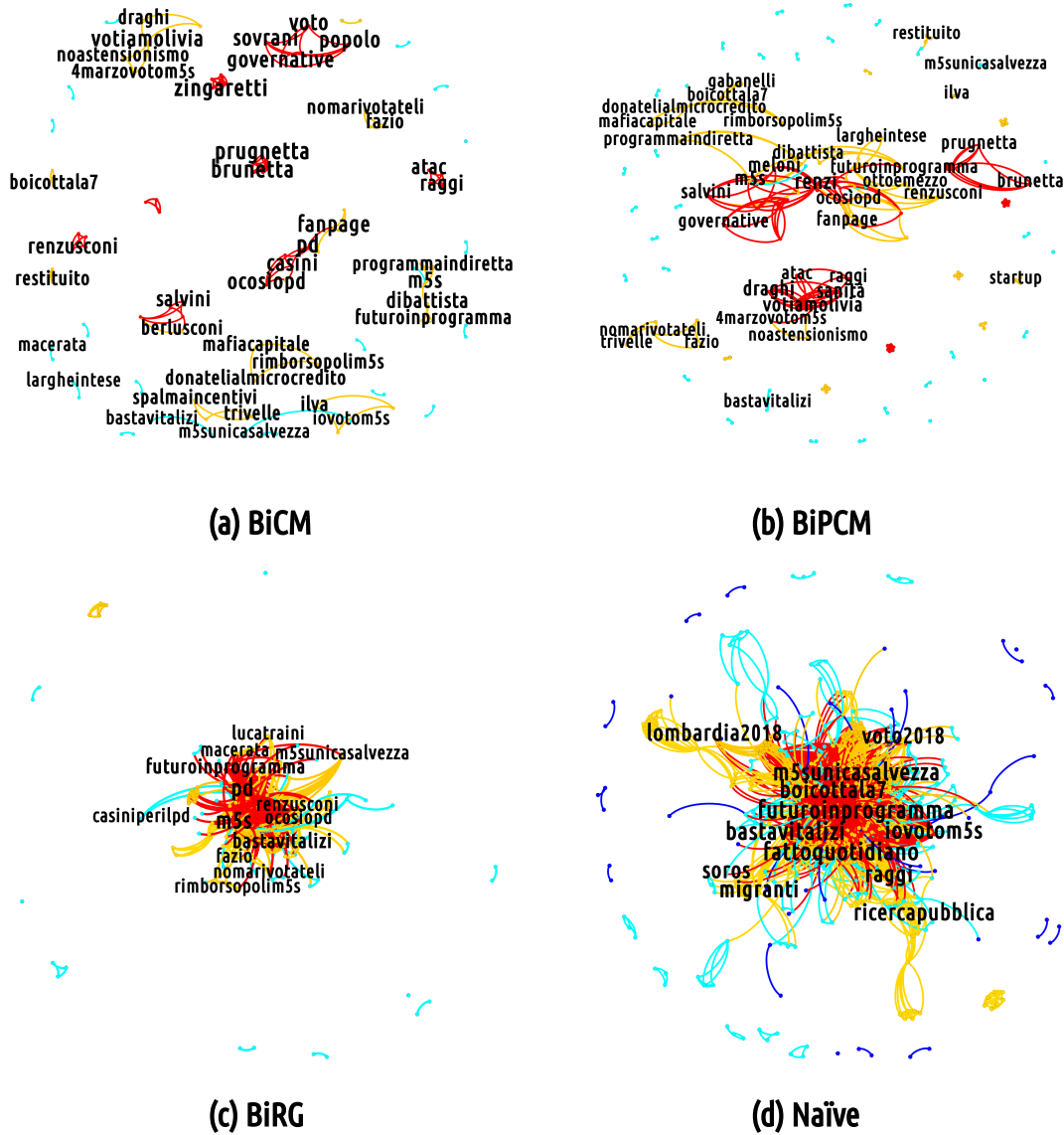}
\caption{Mesoscale structure of (from bottom-right, clockwise) the non-filtered projection of the semantic network corresponding to the M5S discursive community on 19 February 2018 and of the projection of the same network filtered according to the BiRGM, the BiCM and the BiPCM, respectively. The core portion of this network survives the most restrictive filtering (i.e. the BiCM-induced one), indicating that basically \emph{all} hashtags representing topics of interest of the 2018 Italian electoral campaign persist. [Picture realized with Gephi software version 0.9.2. For more information about Gephi \cite{ICWSM09154}: \url{https://gephi.org/}]}
\label{fig:10}
\end{figure}

\begin{figure}[ht!]
\centering
\includegraphics[width=\textwidth]{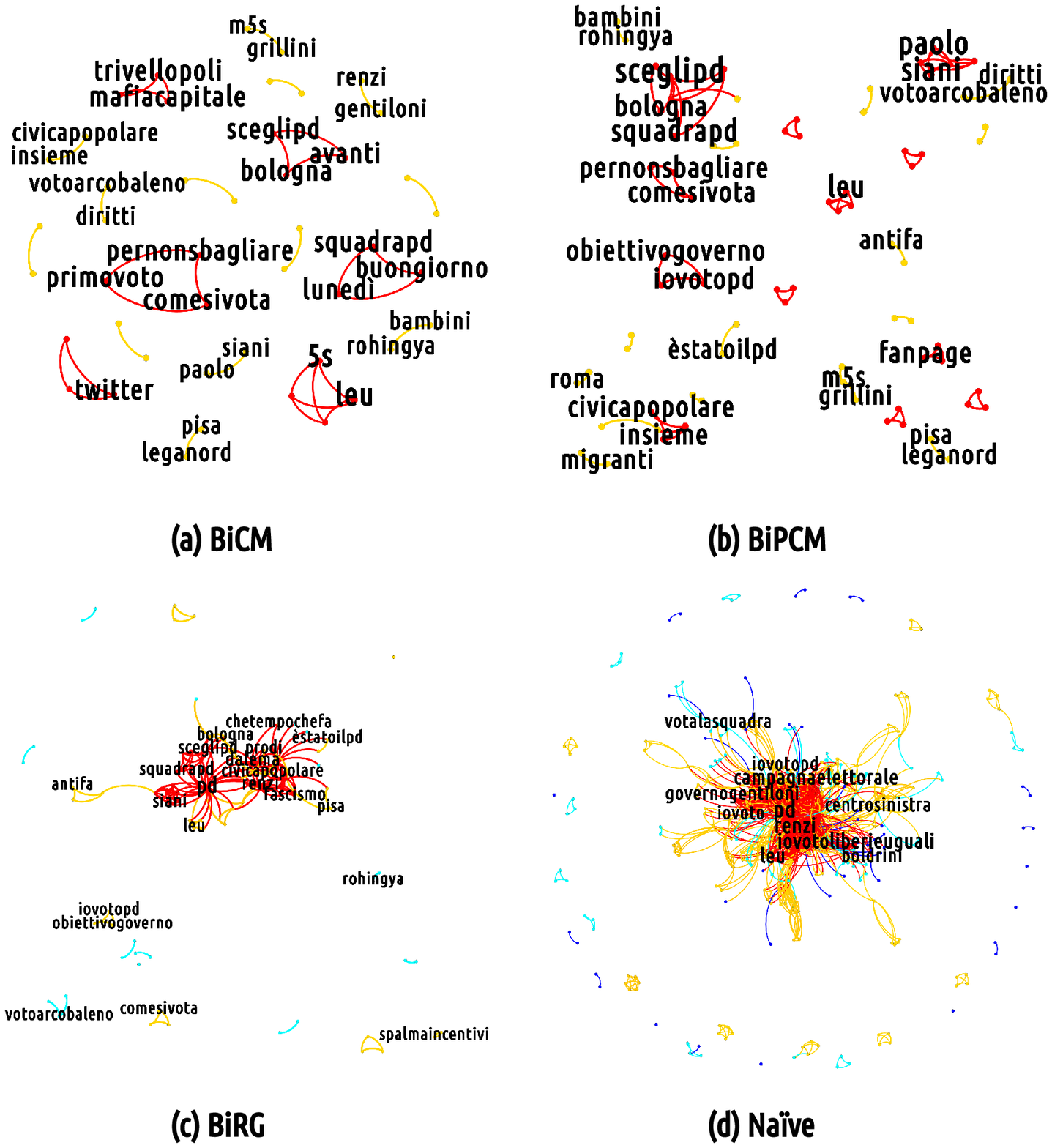}
\caption{Mesoscale structure of (from bottom-right, clockwise) the non-filtered projection of the semantic network corresponding to the CSX discursive community on 19 February 2018 and of the projection of the same network filtered according to the BiRGM, the BiCM and the BiPCM, respectively. The core portion of this network just partially survives the most restrictive filtering (i.e. the BiCM-induced one), while it is present in the less strict filtering (the BiPCM and the BiRGM induced), representing a structure in between the stronge persistence of the M5S semantic network of fig. \ref{fig:10} and the CDX one depicted in \ref{fig:9}. [Picture realized with Gephi software version 0.9.2. For more information about Gephi \cite{ICWSM09154}: \url{https://gephi.org/}]}
\label{fig:11}
\end{figure}

As mentioned above and as fig. \ref{fig:9}, \ref{fig:10}, \ref{fig:11} show, the overall effect of adopting a filtering procedure - regardless of its peculiarities - is that of reducing the total volume of the semantic networks. Differences exist, instead, when it comes to the analysis of nodes' mean degree. Particularly interesting in this respect is the behavior of the semantic networks of the M5S discursive community whose mean degree is affected to a much lesser extent by the BiRGM-induced filtering than those of the CDX and the CSX communities. This, in turn, implies that the information encoded into the total number of retweets of the M5S bipartite user-hashtag network is able to account for the co-occurrences between any two hashtags less effectively than for the CDX and the CSX configurations. Similarly, it is
possible to state that the information encoded into the BiRGM (which is the simplest filter) recognizes the initial structure of the M5S bipartite user-hashtag network as significant.

For what concerns topics persistence, the ranking observed above with reference to the non-filtered projection basically coincides with the ranking observed on the filtered ones. Regarding topics centrality, instead, we observed that the filtering procedure with increasingly restrictive benchmarks makes "emerge" previously screened hashtags (e.g. \textit{\#sicurezza}, \textit{\#fallimentocinquestelle} and \textit{\#precariato}, respectively for the semantic networks induced by the CDX, CSX and M5S discursive communities). Centrality (e.g. the betweenness variant) is, in fact, a highly non-trivial feature that, generally speaking, is not reproduced by the information encoded into the degree sequence alone (not to mention the one encoded into the number of links): in fact, as figs. \ref{fig:9}, \ref{fig:10}, \ref{fig:11} clearly show, only the hashtags belonging to the innermost shells survive our filtering procedure.

Let us now move to discuss the mesoscale structure of the filtered projections looking, as usual, at one specific day that shows the richest structure (again 19 February 2018). Filtering the projections by adopting an increasingly restrictive benchmark make the projection sparser while letting  less trivial structures emerge. Interestingly, the core portion of the semantic network corresponding to the M5S discursive community survives the most restrictive filtering (i.e. the BiCM-induced one), signalling the presence of a non-trivial group of keywords constituting the bulk of the communication in that community (see fig. \ref{fig:9}). Moreover, basically \emph{all} hashtags representing topics of interest of the 2018 Italian electoral campaign persist. The same conclusion holds true for the number of triadic closures observed in correspondence of mediated events: their number is significantly larger with respect to a network model accounting for the total number of tweets only. This result is in line with what has been found for other socio-economic systems (e.g. the World Trade Web~\cite{2015NatSR.510595S}) whose abundance of triadic closures is not reproduced by a benchmark model constraining the total number of links only.

In the following we will describe with more details the main characteristics of the filtered projections of the various semantic networks.

\paragraph{The CDX discursive community.} Figure \ref{fig:9} depicts the semantic network of the center-right alliance on 19 February 2018.

In the BiCM projection, i.e. the strictest one, few links survive. In this situation, it becomes almost inappropriate to talk about communities, since we can find only links connecting two otherwise isolated nodes, or small cliques and chains. Nevertheless, even these few hashtags carry important information regarding the keywords used during the electoral campaign. It is the case of the cluster including \textit{\#stopislam}, \textit{\#stopinvasione} (\emph{stop the invasion}), \textit{\#cdm} (the acronym for the Italian Council of Ministers) and \textit{\#forzeordine} (\emph{law enforcement agencies}), asking for stronger countermeasures to the immigration flows from Northern Africa, perceived as dangerous for the security and for Italian cultural identity. On a similar topic, there is a clique composed by \textit{\#roma}, \textit{\#labaro} and \textit{\#primaporta}: the hashtags refer to neighborhoods in Rome, in which, during the days of the data collection, some thefts in apartments were reported. These hashtags were used to criticise the city administration of Rome, run by Virginia Raggi of the M5S. Moreover, a pair of nodes which represents insulting nicknames for the political opponents are connected between themselves. Those hashtags, \textit{\#pdioti} (i.e., an hashtag mixing the acronym of the \emph{Partito Democratico} and the word \emph{idiots}) and \textit{\#m5stellisti}, are present in a popular message displaying both hashtags, and, in particular, it is the only message displaying \textit{\#m5stellisti}. There is also a clique formed by \textit{\#casapound} (i.e., a neo-fascist party), \textit{\#rai2} (i.e., the second channel of the national TV public service) and \textit{\#19febbraio}. This clique is the result of a viral tweet intended to advertise the presence of the leader of Casa Pound in a public debate held on Rai2 on 19 February. Finally, the last clusters present in the BiCM-induced projection are more institutional: the first contains \textit{\#torniamoagovernare} (\emph{let's go back to govern}), \textit{\#elezioniregionali2018} (\emph{2018 regional administrative election}) and \textit{\#salvini}, while the other one is composed by \textit{\#flattax}, \textit{\#programma} (\emph{program}) and \textit{\#veneto}. The latter set of hashtags is related to an event where the flat taxation government, as part of the electoral program,
is presented.

The BiPCM projection displays a structure in which the various sub-groups described above are reinforced (for instance, the chain \textit{\#flattax}, \textit{\#programma} and \textit{\#veneto} is closed in a clique) and introduces new topics as \textit{\#calenda} (the Minister of the industrial development at the time of the electoral campaign) \textit{\#ilva} and \textit{\#alitalia}, respectively the greatest European steel factory which had severe problems for its environmental, health and economic sustainability, and the Italian national airline, which has been at default risk in the last years. These hashtags are intended to criticize the action of the government in charge at that time. Interestingly, another cluster consisting of the hashtags \textit{\#sullepalle}, \textit{\#fazio}, \textit{\#salvini} is detected by the validated BiPCM projection. These hashtags need a bit of context: during the electoral campaign, the journalist Fabio Fazio invited politicians from all political coalitions to his TV program \textit{Che Tempo Che Fa} broadcast on the national television service, to promote their campaign. Fazio has been accused by all political forces of being too condescending with their opponents. Salvini refused Fazio’s invitation, publicly with insulting language his aversion for the journalist. These hashtags, together with others related to right-wing campaign topics such as \textit{\#vita} and \textit{\#famiglia} (\emph{life} and \emph{family}, related to the Italian anti-abortion movement) are associated with the communication strategy of the most radical part of the center-right alliance.

In the BiRGM validated projection, the clusters found in the previous stricter projections are merged together to form a network organised along two poles: the first is more "institutional" with keywords related to the electoral campaign of \textit{Forza Italia} (the political party of Silvio Berlusconi), including hashtags such as \textit{\#campagnaelettorale} (\emph{electoral campaign}), \textit{\#unitisivince} (\emph{united we will win}), \textit{\#votaforzaitalia} (\emph{vote Forza Italia}); the second is linked to the other two right-wing parties with both the names of their leaders (\textit{\#salvini} and \textit{\#meloni}), but also including their opponents, as \textit{\#pd}, \textit{\#renzi}, \textit{\#pdioti} and \textit{\#m5stellisti}. Interestingly, both poles are organised with a core and a periphery: the two cores are connected by the hashtag \textit{\#centrodestra} (\emph{center-right}), the peripheries by \textit{\#casapound} (i.e., the aforementioned neo-fascist party).

\paragraph{The M5S discursive community.} Overall, the communication strategy of the M5S is peculiar since the users tend to use a quite large number of hashtags and thus discuss a higher number of topics. Hashtags are nearly the same across various tweets, since they were copied and pasted from older messages on the same topics. In this sense, Twitter users in the M5S community appear to be more coordinated and thus manage to give their hashtags more visibility.

Considering the tweets and retweets published on 19 February, the M5S validated semantic networks of fig. \ref{fig:10} displays a rich structure, even in the BiCM projection, due to the tweeting behavior described above. In particular, several clusters can be found, including the name of the opponents (\textit{\#renzi}, \textit{\#salvini}, \textit{\#gentiloni}, \textit{\#pd}), or few nicknames assigned to them (as \textit{\#prugnetta}, \emph{little plum}, for Brunetta, a member of \textit{Forza Italia}; \textit{\#renzusconi}, a mix between Renzi and Berlusconi, intending that there is few differences between them) or other slogans teasing political opponents (\textit{\#votiamolivia}, \emph{let's vote them away}; \textit{\#nomarivotateli}, \emph{no, but vote them again}, ironically targeting PD supporters; \textit{\#ocos\`iopd}, \emph{this way or PD's way}, advertising that
the only political alternative to \textit{Partito Democratico} is the \textit{Movimento Cinque Stelle}). Few clusters represent some events in the electoral campaign. For instance, a cluster following the electoral campaign tour of Di Battista, a member of the \textit{Movimento Cinque Stelle}, appears in this projection. Even a clique advertising a live streaming on Facebook can be observed, discussing the management of the public health system in the Lazio region (governed by the PD), with the hashtags \textit{\#lazio}, \textit{\#sanit\`a} (\emph{public health system}), \textit{\#sancesareo} (the town were the live streaming was set), \textit{\#zingaretti} (the president of the Lazio region).

The topic of bad governance of the political opponents represents a big part of the semantic network of the M5S community: in addition to the cluster mentioned above, another cluster focuses on the news about a journalist who was attacked during a campaign event held by the center-left coalition in Naples (\textit{\#fanpage} which is the online newspaper for which the journalist worked; \textit{\#inchiestanapoli}, \emph{Naples investigation}; \textit{\#video}). Moreover, the hashtags \textit{\#donatelialmicrocredito} (\emph{give them to the microcredit}) and \textit{\#rimborsopoli} (\emph{refund scandal}) refer to the scandal of a criminal organisation in Rome bribing members of established political parties. The \textit{Movimento Cinque Stelle} expelled its representatives involved in this investigation and proposed that other parties do the same and transfer the amount of the bribe to
the microcredit. There are also clusters targetting harsh political debates, such as the case of \textit{\#dibiase}, referring to Letizia Di Biase who is the wife of the Italian Minister of Cultural Heritage and Activities. After being elected member of the council in the city of Rome, she
did not resign when elected as member of the council of the region of Lazio. Di Biase was also criticised for criticizing the mayor of Rome Virginia Raggi of the \textit{Movimento Cinque Stelle} for filing for bankruptcy for the city agency of public transportation while salvaging the regional one operated by the regional administration of the \textit{Partito Democratico}. Finally, there are traces of the debate between the virologist Roberto Burioni and the Head of the Italian Order of Biologists, Vincenzo D’Anna, concerning the presence of anti-vaccine groups and \textit{Movimento Cinque Stelle} supporters during a national conference of the order of biologists. Other hashtags refer to Giorgia Meloni (leader of \textit{Fratelli d'Italia}) and the charges moved against her for leaning close to neo-fascist parties and ideology.

The BiPCM projection increases the connections among the topics and few hashtags appear to be related to names of places covered by the campaign tour of Carlo Sibilia, another exponent of the M5S. Instead, the BiRGM projection displays a strong core-periphery structure.

\paragraph{The CSX discursive community.} In the case of the CSX community, the number of hashtags used is relatively small. With respect to other discursive communities, the semantic network of the center-left group validated by the BiCM (see fig. \ref{fig:11}) focuses more on political subjects, as shown by the pair \textit{\#diritti} (\emph{rights}) and \textit{\#arcobaleno} (\emph{rainbow}; both hashtags refer to the LGBTQIA+ civil rights; but also by the coupling of  \textit{\#bimbi} (\emph{children}) and  \textit{\#rohingya} (related to the subject of the Rohingya exodus in Myanmar and the condition of young children). Other clusters pivot around instructions for youngsters voting for the first time (\textit{\#primovoto}, \emph{first vote}; \textit{\#comesivota}, \emph{how to vote}; \textit{\#pernonsbagliare}, \emph{how to avoid mistakes}) and call for fact checking during the electoral campaign, with hashtags \textit{\#factchecking} and \textit{\#checkpolitiche2018}.

Interestingly, a clique linked to the particular virality of one specific tweet, is formed by the hashtags \textit{\#trivellopoli}, \textit{\#mafiacapitale} and \textit{\#consip}, i.e. three scandals in which the \textit{Partito Democratico} was involved. This tweet suggests that those scandals suspiciously appeared during the electoral campaign in order to damage the name of the \textit{Partito Democratico} thereby limiting its performance at the general elections.

Another conversational line unfolds along the candidacy of Paolo Siani, a physician particularly active in providing support, in collaboration with local NGOs, to children of the poor neighborhoods of Naples at risk of being recruited for organized criminal activities. More broadly, the public presentation of the \textit{Partito Democratico} candidates team constitutes a topic widely debated within this discursive
community as shown by the two hashtags \textit{\#renzi} and \textit{\#gentiloni}, respectively the National Secretary of the \textit{Partito Democratico} and the candidate for the position of Prime Minister, in connection with other hashtags. For instance, the clique \textit{\#bologna}, \textit{\#avanti} (\emph{let's move forward}) and \textit{\#sceglipd} (\emph{choose PD}) refers to an event led by Paolo Gen-
tiloni and Matteo Renzi in Bologna while the clique \#luned\`i (\emph{Monday}), \textit{\#buongiorno} (\emph{good morning}) and \textit{\#squadrapd} (\emph{PD team}) appeared in a message promoting a massive electoral campaign. 

In the BiPCM validated projection more connections appear, enriching topics which had already emerged, as in the case of the candidacy of Paolo Siani mentioned above: \textit{\#babygang}, \textit{\#napoli} (\emph{Naples}), \textit{\#infanzia} (\emph{childhood}) merge with the previous hashtags \textit{\#paolo} and \textit{\#siani}. A new cluster containing the name of the opponents (\textit{\#dimaio}, \textit{\#salvini}, \textit{\#meloni}, \textit{\#fascismo}) is also present.

In the BiRGM validated projection, the aforementioned structures gain new links and new nodes and a richer structure becomes evident. In particular, three main communities appear: one (in orange in fig. \ref{fig:11}) pivoting around political adversaries (including \textit{\#salvini}, \textit{\#meloni}, \textit{\#dimaio}, \textit{\#grillini}, \textit{\#berlusconi}), one advertising political subjects and events of the electoral campaign (including \textit{\#sceglipd}, \emph{choose PD}; \textit{\#squadrapd}, \emph{PD team}; \textit{\#diritti}, \emph{rights}, and so on) and one related to the candidacy of Paolo Siani. A peripheral clique advertising the event in Venice of \emph{Liberi e Uguali}, a political party on the left of \textit{Partito Democratico}, can also be found (\textit{\#antifa}, \textit{\#liberieuguali}, \textit{\#venezia}, \emph{Venice}).

\paragraph{Final remarks on the filtering procedure.} Summing up, in all na\"ive projections we observe a rich structure, with a particularly evident core-periphery organisation. This structure is progressively disintegrated through filtering and depending on the strictness of the benchmark used. While this disintegration affects semantic structures generated by all discursive communities, the various groups display a different resilience to the filtering procedure: in particular, the one revolving around M5S accounts are the one with the least trivial structure, hence being affected the least. Quite relevantly, the different network structures carry information about the strategy followed by the various discursive communities during their political campaign.

The validation procedure proposed in \cite{Saracco_2017} projects non-trivial co-occurrences of links in the bipartite networks, i.e. those that are not explained by the ingredients of the null-model used for filtering. In this sense, validated nodes in the projection are not necessarily those with, for example, the highest (bipartite) degree, but those grappling to other hashtags in the semantic network to a larger extent than what expected by just looking at the original bipartite network. With respect to the examined case study, the validated projections show that the more the validated links, the more hashtags are used to refer to a single subject, against the random superposition of ubiquitous slogans. This seems to be the case particularly for the M5S community while it is true to a much lesser extent for the CDX discursive community where the amount of nodes in the BiCM validated projection is extremely limited. The validation procedure allows us to focus on the least trivial connections and thus to observe different conversational lines that shape the political communication of the various discursive communities. In this way, the validation procedure allows to uncover otherwise invisible conversational lines that shape the Twitter activity of each discursive community.

In the CDX, a clear thematic distance is present between the far right (formed by parties led by Matteo Salvini and Giorgia Meloni) and center-right politicians (Silvio Berlusconi and his party \emph{Forza Italia}) in terms of topics and electoral slogans. While the former insists on security issues related to migration flows from Northern Africa, the latter tends to promote a united center-right alliance. There is an evident semantic diversification with completely different keywords used in the tweets: the far right uses more aggressive statements and bad words, while the second is more reassuring and institutional. \\

The M5S projected semantic networks are especially rich in structure, due to the strong usage of hashtags in this community. Most of them are referring to political opponents with nicknames and ironic slogans. A great part of the filtered semantic network is devoted to highlight the deceitfulness of the M5S opponents.\\ 

Validated semantic networks of the CSX community are poorer than those of M5S, but richer than those of the CDX. Their major feature is to present mostly events of the electoral campaign, their candidates at a national and regional level and the weaknesses of their political opponents. 

It is worth noticing that the peculiarities of the three discursive filtered semantic networks are present in other days which are not explicitly commented here. For instance, on 11 February, we can still observe two different poles of the debates in the CDX community, one promoted by the supporters of \emph{Forza Italia} and the other promoted by the supporters of far-right wing parties. As observed for the 19 of February, the two poles use different vocabularies and focus, respectively, on reforming taxation and labour or on the migration issues. Analogously, the M5S displays a cluster of discussion against the use of vaccines, few clusters against an alleged \emph{quid pro quo} between some PD members and some businessmen as well as some other cluster teasing political opponents. Finally, the CSX focus on the election candidates presentation and few national problems (increasing inequality, poverty and the decreasing birth-rate). In all three networks there are also mentions to the demonstration involving nearly thirty-thousand persons against neo-fascism held in Macerata on 10 February, albeit with different levels of attention. 


\section*{Discussion}\label{sec:discussion}


In this section, we would like to discuss more in details some relevant features of our methodological approach and to summarize results about the Italian case we examined throughout this work. Our study focused on the activity on Twitter of three different discursive communities tied to the main political topics competing during general elections in Italy in 2018. It exploited approximately 1 million of tweets which we used to define networks of statistically significant co-occurrences of hashtags at a daily time scale. Compared to extant research, we argue that our proposed approach innovates studies in this domain in two main ways. First, we propose a methodological framework \cite{Becatti2019c,2021arXiv210304653R} to investigate the semantic components of online political discussions coupling attention for topics prominence and persistence with a sound exploration of how they are organized and structured along conversational lines. Second, we did not treat the semantic aspect in isolation but, in fact, built a tight connection between contents discussed by users and the type of interactions they performed via Twitter. Moreover, our way of isolating discursive communities and related semantic networks does not entail any manual intervention. Initial distinction between verified and non-verified accounts is not defined by researchers but is, in fact, assigned by the Twitter platform itself. Relevantly, as the categorization of accounts according to these two categories is regularly provided through the Streaming API by the platform itself, the solidity of the initial partition is ensured platform-wise. Similarly, topics and conversational lines within semantic networks are not biased by construction but, in fact, are a direct output of the application of filtering procedures with maximally random benchmarks.

Against this background, we sought to offer a solid analysis of the public discussion developed online during a crucial political event and to infer, starting from semantic aspects, marking traits of the different political identities that guide political courses of action in the Italian political scenario. One of the main findings of this paper concerns the way the topological structure of semantic networks "reacts" to the so-called mediated events (i.e. TV debates, the media coverage of offline events, etc.) thus revealing not only different sensibility towards the media sphere but, more significantly, different identity traits that characterize each discursive political community. The topology of the CDX community is strongly dependent on these events (the mean degree of nodes increases in correspondence of specific TV shows), meaning that this group of users is more involved in the activity of retweeting during the appearance of political actors particularly on television. Conversely, the activity of the M5S community appears to be much more "distributed". In fact, although M5S supporters are sensitive to TV shows as well, their retweeting activity is not exclusively driven by media events but, rather, follows their preference for a generalized use of social media for organizational and political communication activities. Finally, the activity of the CSX community is characterized by a somewhat "intermediate" behavior: even when mediated events affect the Twitter discussion, the attention of the whole community is somehow "shared" among the various actors constituting the center-left alliance.

Particularly insightful is the analysis of our semantic networks at the mesoscale: what emerges is the presence of a core of topics, i.e. a densely-connected bulk of hashtags surrounded by a periphery of loosely inter-connected (sub-)topics. This indicates that daily semantic networks are characterized by few relevant hashtags to which other, less relevant ones, attach. This structure is maintained during the whole observation period and differences emerge only with respect to the number of \emph{peripheral} themes entering the discussion. The resilience of the core-periphery structure is not the same for the various discursive communities. In the context of semantic networks, the fact that the system is more or less resilient to the filtering implies that the various political groups have developed differently their political narrative, focusing their communications on few related terms per subject or mentioning a set of omnipresent hashtags in all messages. Even in the response to the filtering procedure, the M5S and the CDX communities represent the two extremes, displaying respectively the most and the least resilient semantic network.

These differences are the effect of various styles used in writing posts. When targeting a specific theme, members of the M5S community use several hashtags that are subsequently re-used by other users writing on the same argument in order to make keywords and slogans more recognizable and visible. Conversely, in the CDX community, the number of hashtags per message is more limited and users tend to use particularly viral hashtags. Moreover, the CDX shows a diversified communication strategy, hence a possible internal political fracture, due to the different approaches of the various parties in the alliance: right-wing politicians are more aggressive towards opponents, while center-right ones tend to focus on unitive (for the coalition) keywords. Somewhere in between, the CSX community balances a use of hashtags to criticize its adversaries but also to amplify the reach of events and initiatives of the electoral campaign.


\section*{Conclusions}\label{sec:conclusions}

Social media platforms have dramatically changed patterns of news consumption and, over the last years, they have become increasingly central during political events, especially electoral campaigns. In this respect, Twitter has been shown to play a major role, thus attracting the attention of scientists from all disciplines. So far, however, researchers have mainly focused on users' activity, paying little attention to semantic aspects which are instead particularly relevant to detect online debates, understand their evolution and, ultimately, inferring the behavioral rules driving (online but also offline) electoral campaigns.

As claimed in the Section \hyperref[sec:introduction]{Introduction}, our goal in this paper is to advance a comprehensive methodological framework to fill a persistent attention gap for the semantic aspects of online political discussions. Accordingly, we did put forward a scheme of analysis that couples attention for more evident aspects, such as topic visibility, persistence and strategic uses in conversation, with the systematic analysis of more invisible, and yet crucial aspects of content production and circulation - particularly, the identification of conversational cores and main lines of development. In doing so, we believe that our proposed approach provides a solid starting point to understand the symbolic aspects that flank and, in fact, nurture current dynamics of online civic participation, political partisanship and polarization, which have so far catalyzed attention within the research community.

More relevantly, our proposed approach did not bring prominence to semantic aspects leaving the social side of online dynamics behind. The semantic networks we analyzed in this paper follows from an identification of discursive communities that leans on an entropy-based framework, which is a methodological advancement \textit{per se} \cite{Becatti2019c,2021arXiv210304653R}. On the one hand, this technique allowed us to filter the retweeting activity of users while singling out the statistically-relevant information at the desired level of detail. Consistently, the semantic structures induced from communities are indicative of users’ political affiliation yet without passing through any manual labelling of our media contents (e.g. as performed in \cite{10.5555/3382225.3382281,6113114}). On the other hand, we deduce the discussions taking place in each community by connecting any two hashtags if used a significantly large number of times by the users of that community, hence overcoming the limitations of other analyses \cite{doi:10.1177/2056305117704408,XIONG201910} where online political conversations are studied in an unrelated fashion with respect to the relational system amongst users sustaining them.

Our proposed approach allowed to reach several fine-grained insights about the Italian case both at the macro level, grasping the semantic peculiarities of broader conversations taking place within discursive communities, and, at the micro level, narrowing down our exploration to single and meaningful points in time during the electoral campaign period. Above and beyond the particular case study we analysed, we believe that the same approach can be applied to disentangle the intricacies of large-scale Twitter conversations in all domains and regardless of the language in which they take place with the only requirement of extracting the data set with a meaningful list of anchor hashtags.

Paralleling these advantages, our approach remains limited mainly in two respects. First however detailed and multilevel, our semantic analysis remains a partial investigation of the symbolic universe that is produced and circulated online in conjunction with relevant political events and dynamics. The semantic structures we investigate in this paper are formed in the space created by a single platform and pivot around the use of a specific feature for marking contents - i.e., hashtags. Ad-hoc publics that assemble around topics, in fact, are not exhausted by communities that form on particular social media platforms - let alone around specific hashtags \cite{quteprints91812}. Moreover, looking at conversations forming around specific hashtags fails to include those contributions that, albeit pertinent, are delivered without including any specific content markers \cite{hanna2013computer}. The very procedure to discount all those messages that did not contain an hashtag from the initial corpus of tweets containing election-related keywords well illustrates the need to account for multiple communicative actions performed by users.

Second, albeit automatically inferred and not depending in any way from aprioristic assumptions on users' political orientation, our discursive communities are neither representative of the Italian voting population nor an exact indication of users' political affiliation. On the one side, it is widely recognized that Twitter users are not representative of broader populations (in this case, of all Italian citizens) and that Twitter Search API does not guarantee the representativeness of the data themselves \cite{pfeffer2018tampering}. On the other, the identification of discursive communities starts from retweets which, as mentioned above, simply express an explicit bestowal of attention. The total lack of information about the actual political affiliation of users does severely limit our capacity to predict, starting from semantic data, relevant political outcomes such as voters turnout, election results, or the construction of political alliances between parties.

Nonetheless, without any claim of exhaustivity, our mapping of the Twitter discussion in occasion of national elections provides a useful entry point to reason around the online construction of political collective identities. A plethora of studies based on Twitter freely available data has shown that it is indeed possible to infer the political orientation of users from tweets and to analyze electoral debates and societal discussions \cite{Caldarelli2020,Becatti2019,doi:10.1177/0002764213479371,6113114,2021arXiv210304653R,Pavan2019} shedding light on the political implications of non-traditional political acts such as expressing publicly on social media. Users employing in their tweets election-related keywords and hashtags did in fact contribute to frame the electoral campaign period along certain lines and they did so upon a platform that was not only widely diffused in Italy in that specific moment (According to Audiweb \cite{Vincos}, 9 millions of Italian users were active on Twitter in 2018) but that also plays a pivotal political communication role \cite{doi:10.1080/17512786.2015.1040051}. Moreover, the strict filtering procedure we applied leads to the identification of networks with only statistically-relevant hashtags information, as the projections that we adopted guarantees that the analysis of induced semantic networks is sound from both a methodological and interpretative point of view. Thus, albeit non representative of and non generalizable to the overall Italian population, both discursive communities and induced semantic networks that we examined on can be thought as a solid starting ground to develop more fine-grained studies of voters’ political opinion and behaviors.

\bibliographystyle{unsrt}
\bibliography{References}



\section*{Acknowledgements}

F. S. and T.S. acknowledge support from the European Project SoBigData++ (GA. 871042). F.S. also acknowledges support from the Italian `Programma di Attivit\`a Integrata' (PAI) project `TOol for Fighting FakEs' (TOFFE), funded by IMT School for Advanced Studies Lucca. E.P. acknowledges support from the project "I-Polhys - Investigating Polarization in Hybrid Media Systems" funded by the Italian Ministry of University and Research within the PRIN 2017 framework (Research Projects of Relevant National Interest for the year 2017; project code: 20175HFEB3). 

\section*{Author contributions statement}

T.R., E.P, T.S. and F.S. contributed equally to outline the research, interpret the results, write and review the manuscript. T.R., T.S. and F.S. provided the analytical and numerical tools and T.R. performed the analysis.

\section*{Additional information}

The authors declare no competing interests. T.R. is responsible for submitting on behalf of all authors of the paper.

\newpage

\section*{Appendix}

\subsection*{Supplementary Note 1: Defining a similarity measure}

A \textit{sequence-based similarity} quantifies the cost of transforming a string $x$ into a string $y$ when the two strings are viewed as sequences of characters. String transformation is defined by three elementary operations: 1) deleting a character, 2) inserting a character and 3) substituting one character with another \cite{zbMATH03240929}. The edit distance function $d(x,y)$ aims at capturing the mistakes of human editing, such as inserting extra characters or swapping any two characters. To merge only strings that are either misspelled or different by number (i.e. singular in place of plural and viceversa) we have set the threshold for the maximum number of allowed differences between any two strings to 2.

\subsection*{Supplementary Note 2: Projecting and validating bipartite networks}

As anticipated in the main text, the idea behind a filtered projection is that of \emph{linking any two nodes belonging to the same layer if found to be sufficiently similar}. The steps to implement such a procedure are described below.

\paragraph{Quantifying nodes similarity.} First, a measure quantifying the similarity between nodes is needed. Given any two nodes (say, $\alpha$ and $\beta$) we follow \cite{Saracco_2017} and count the total number of common neighbors $V_{\alpha\beta}^*$, i.e.

\begin{equation}
\label{eq:simil}
V_{\alpha\beta}^*=\sum_{j=1}^{N_\top}m_{\alpha j}m_{\beta j}=\sum_{j=1}^{N_\top} V_{\alpha\beta}^j
\end{equation} 
the value of $V_{\alpha\beta}^j$ being 1 if nodes $\alpha$ and $\beta$ share the node $i$ as a common neighbor and 0 otherwise. Notice that the non-filtered projection of a  bipartite network corresponds to a monopartite network (say, $\mathbf{A}$) whose generic entry reads $a_{\alpha\beta}=\Theta[V_{\alpha\beta}^*]$ (i.e. it is an edge in correspondence of any non-zero value of $V_{\alpha\beta}^*$).

\paragraph{Quantifying the statistical significance of nodes similarity.} The statistical significance of any two nodes similarity is quantified with respect to a bunch of null models which will be now derived from first principles. To this aim, let us consider the maximization of Shannon entropy

\begin{equation}
\label{eq:entropy}
S=-\sum_{\mathbf{G}\in\mathcal{G}} P(\mathbf{G})\ln P(\mathbf{G})
\end{equation}
over the set of all, possible, bipartite graphs with, respectively, $N_\top$ nodes on one layer (say, users) and $N_\bot$ nodes on the other (say, hashtags). Since entropy-maximization will be carried out in a constrained framework, let us discuss each set of constraints separately.

\textit{Bipartite Configuration Model.} The \textit{Bipartite Configuration Model} (BiCM) represents the bipartite variant of the Configuration Model (CM). Upon introducing the Lagrangian multipliers $\boldsymbol{\theta}$ and $\boldsymbol{\eta}$ to enforce the proper constraints (i.e. the ensemble average of the degrees of users and hashtags, respectively $h_i^*=\sum_\alpha m_{i\alpha},\:\forall\:i$ and $k_\alpha^*=\sum_i m_{i\alpha},\:\forall\:\alpha$) and $\psi$ to enforce the normalization of the probability, the recipe prescribes to maximize the function

\begin{equation}
\mathcal{L}=S-\psi\left[1-\sum_{\mathbf{G}\in\mathcal{G}}P(\mathbf{G})\right]-\sum_{i=1}^{N_\top}\theta_i[h_i^*-\langle h_i\rangle]-\sum_{\alpha=1}^{N_\bot}\eta_\alpha[k_\alpha^*-\langle k_\alpha\rangle]
\end{equation}
(with respect to $P(\mathbf{G})$. This leads to

\begin{equation}
P(\mathbf{G}|\boldsymbol{\theta},\boldsymbol{\eta})=\frac{e^{-H(\mathbf{G})}}{Z}=\prod_{i=1}^{N_\top}\prod_{\alpha=1}^{N_\bot}\left(\frac{x_iy_\alpha}{1+x_iy_\alpha}\right)^{m_{i\alpha}}\left(\frac{1}{1+x_iy_\alpha}\right)^{1-m_{i\alpha}}=\prod_{i=1}^{N_\top}\prod_{\alpha=1}^{N_\bot}p_{i\alpha}^{m_{i\alpha}}(1-p_{i\alpha})^{1-m_{i\alpha}}
\end{equation}
where $x_i\equiv e^{-\theta_i}$ and $y_\alpha\equiv e^{-\eta_\alpha}$. The quantity $p_{i\alpha}=\frac{x_iy_\alpha}{1+x_iy_\alpha}$ can be interpreted as the probability that a link connecting nodes $i$ and $\alpha$ is there; the matrix of probability coefficients $\{p_{i\alpha}\}$ induces the expected values $\langle h_i\rangle=\sum_\alpha p_{i\alpha},\:\forall\:i$ and $\langle k_\alpha\rangle=\sum_i m_{i\alpha},\:\forall\:\alpha$ and can be numerically determined by solving the set of $N_\top+N_\bot$ equations $\langle h_i\rangle=h_i^*,\:\forall\:i$ and $\langle k_\alpha\rangle=k_\alpha^*,\:\forall\:\alpha$.

According to the BiCM, the presence of each $V_{\alpha\beta}^j$ can be described as the outcome of a Bernoulli trial:

\begin{eqnarray}
f_\text{Ber}(V_{\alpha\beta}^j=1)&=&p_{\alpha j}p_{\beta j}, \\
f_\text{Ber}(V_{\alpha\beta}^j=0)&=&1-p_{\alpha j}p_{\beta j}.
\end{eqnarray}

The independence of links implies that each $V_{\alpha\beta}$ is the sum of independent Bernoulli trials, each one characterized by a different probability. The behavior of such a random variable is described by a Probability Mass Function (PMF) called Poisson-Binomial.

\textit{Bipartite Partial Configuration Model.} The BiCM constrains the degrees of both the users and the hashtags. Such a model can be `relaxed' by limiting ourselves to constrain the degrees of the nodes belonging to the layer of interest - in this case, the degrees of the hashtags. Upon `switching off' the user-specific constraints, one end up with a simplified version of the BiCM, characterized by a generic probability coefficient reading $p_{i\alpha}=\frac{h_\alpha^*}{N_\top}$, in turn leading to the expression $f_\text{Ber}(V_{\alpha\beta}^j=1)=\frac{h_\alpha^*h_\beta^*}{N_\top^2}$. The evidence that the latter expression does not depend on $j$ simplifies the description of the random variable $V_{\alpha\beta}$, now obeying a PMF called Binomial, i.e.

\begin{equation}
f_\text{BiPCM}(V_{\alpha\beta}=n)=\binom{N_\top}{n}\left(\frac{h_\alpha^*h_\beta^*}{N_\top^2}\right)^n\left(1-\frac{h_\alpha^*h_\beta^*}{N_\top^2}\right)^{N_\top-n}.
\end{equation}

\textit{Bipartite Random Graph Model.} The BiRGM (Bipartite Random Graph Model) is the bipartite variant of the traditional Random Graph Model. As for its monopartite counterpart, the probability that any two nodes are linked is equal for all the nodes and reads $p_{i\alpha}=\frac{N_{\top}N_{\bot}}{L}\equiv p_\text{BiRGM}$ (where $L$ is the empirical number of `bipartite' edges). In this case, we have $f_\text{Ber}(V_{\alpha\beta}^j=1)=p_\text{BiRGM}^2$ and the PMF describing the behavior of $V_{\alpha\beta}$ is a Binomial, i.e.

\begin{equation}
f_\text{BiRGM}(V_{\alpha\beta}=n)=\binom{N_\top}{n}(p_\text{BiRGM}^2)^n(1-p_\text{BiRGM}^2)^{N_\top-n}.
\end{equation}

\paragraph{Validating the monopartite projection.} The statistical significance of the similarity of nodes $\alpha$ and $\beta$, thus, amounts at computing a p-value on one of the aforementioned probability distributions, i.e. the probability of observing a number of V-motifs greater than, or equal to, the observed one:

\begin{equation}
\label{eq:p_value}
\mbox{p-value}(V_{\alpha\beta}^{*})=\sum_{V_{\alpha\beta}\geq V_{\alpha\beta}^{*}}f(V_{\alpha\beta}).
\end{equation}

After this procedure is repeated for each pair of nodes, an $N_\bot\times N_\bot$ matrix of p-values is obtained. The choice of which p-values to retain has to undergo a validation procedure for testing multiple hypotheses at the same time: here, the False Discovery Rate (FDR) procedure is used. The $m$ p-values (in our case, $m=N_\bot(N_\bot-1)/2$) are, first, sorted in increasing order, p-value$_{1}\leq\ldots\leq$p-value$_m$ and, then, the largest integer $\hat{i}$ satisfying the condition

\begin{equation}
\mbox{p-value}_{\hat{i}}\leq\frac{\hat{i}t}{m}
\end{equation}
(where $t$ represents the single-test significance level - in our case, set to 0.05) is individuated. All p-values that are less than, or equal to, $\mbox{p-value}_{\hat{i}}$ are kept, i.e. all node pairs corresponding to those p-values will be linked in the resulting monopartite projection.

\subsection*{Supplementary Note 3: Analysing a network mesoscale structure}

\subsubsection*{Community detection: the Louvain algorithm}
\label{sec:communitydetection}

After the daily monopartite user networks have been obtained, the Louvain algorithm \cite{Blondel_2008} has been run to detect the presence of communities. This algorithm works by searching for the partition attaining the maximum value of the modularity function $Q$, i.e.

\begin{equation}
Q=\frac{1}{2L}\sum_{i,j}\left[a_{ij}-\frac{k_ik_j}{2L}\right]\delta_{c_i,c_j}
\end{equation}

a score function measuring the optimality of a given partition by comparing the empirical pattern of interconnections with the one predicted by a properly-defined benchmark model. In the expression above, $a_{ij}$ is the generic entry of the network adjacency matrix $\mathbf{A}$, the factor $\frac{k_ik_j}{2L}$ is the probability that nodes $i$ and $j$ establish a connection according to the Chung-Lu model, $\bm{c}$ is the $N$-dimensional vector encoding the information carried by a given partition (the $i$-th component, $c_i$, denotes the module to which node $i$ is assigned) and the Kronecker delta $\delta_{c_i,c_j}$ ensures that only the nodes within the same modules provide a positive contribution to the sum. The normalization factor $2L$ guarantees that $-\frac{1}{4}\leq Q(\bm{c})\leq1$. Moreover, a reshuffling procedure has been applied to overcome the dependence of the original algorithm on the order of the nodes taken as input.

\subsubsection*{Core-periphery detection}
\label{sec:cpdetection}

Core-periphery detection can be carried out upon adopting the method proposed in~\cite{de2019detecting} and prescribing to search for the network partition minimizing the quantity called \emph{bimodular surprise}, i.e. 

\begin{equation}
\mathscr{S}_\parallel=\sum_{i\geq l_\bullet^*}\sum_{j\geq l_\circ^*}\frac{\binom{V_\bullet}{i}\binom{V_\circ}{j}\binom{V-(V_\bullet+V_\circ)}{L-(i+j)}}{\binom{V}{L}};
\label{Core_periphery_eq}
\end{equation}
as anticipated in the main text, $L$ is the total number of links, while $V$ is the total number of possible links, i.e. $V=\frac{N(N-1)}{2}$. The quantities marked with $\bullet$ ($\circ$) refer to the corresponding core (periphery) quantities, i.e. $V_\bullet$ is the total number of possible core links, $V_\circ$ is the total number of possible periphery links, $l_\bullet^*$ is the number of observed links within the core and $l_\circ^*$ is the number of observed links within the periphery.

From a technical point of view, $\mathscr{S}_\parallel$ is the p-value of a multivariate hypergeometric distribution, describing the probability of $i+j$ successes in $L$ draws (without replacement), from a finite population of size $V$ that contains exactly $V_\bullet$ objects with a first specific feature and $V_\circ$ objects with a second specific feature, wherein each draw is either a `success' or a `failure': analogously to the univariate case, $i+j\in [l_\bullet^*+l_\circ^*,\min\{L,V_\bullet+V_\circ\}]$. The method outputs the most statistically significant core-periphery structure compatible with the network under analysis.

\subsection*{Supplementary Note 4: Computing the polarization of non-verified users}
\label{sec:userpolarization}

Let $C_{c}$, with $c=1,2,3$, indicate the set of (both verified and non-verified) users belonging to community $c$ and $N_{\alpha}$, with $\alpha=1,2,3$ the set of neighbours of verified users belonging to the community $c=\alpha$. A non-verified user \emph{polarization} is defined as

\begin{equation}
\rho_{\alpha}=\max_c\{I_{\alpha c}\}
\end{equation}
where

\begin{equation}
I_{\alpha c}=\frac{|C_c\cap N_\alpha|}{|N_\alpha|}.
\end{equation}

As it has been shown in \cite{Becatti2019}, the polarization index reveals how unbalanced is the distribution of interactions between non-verified users and verified users: non-verified accounts basically focus their retweeting activity on the tweets of verified users within the same community, thus providing a clear indication of the community of which a non-verified user is likely to be a member.

\end{document}